\documentclass[aps,pra,reprint,superscriptaddress]{revtex4-2}

\usepackage{amsmath,amssymb,mathtools,bm}
\usepackage{physics}
\usepackage{braket}
\usepackage{graphicx}
\usepackage{booktabs}
\usepackage{multirow}
\usepackage{color}
\usepackage{tikz}
\usetikzlibrary{arrows.meta,calc,positioning}
\usepackage[colorlinks=true,linkcolor=blue,citecolor=blue,urlcolor=blue]{hyperref}

\newcommand{\calF}{\mathcal{F}}

\newcommand{\etaT}{\eta_{T}}
\newcommand{\etaG}{\eta_{G}}
\newcommand{\etaR}{\eta_{R}}
\newcommand{\RP}{R_{P}}
\newcommand{\RE}{R_{E}}
\newcommand{\PBP}{PBP}
\newcommand{\allBSM}{all-BSM}
\newcommand{\Rphi}{R_{\phi}}
\newcommand{\Uphi}{U_{\phi}}

\newcommand{\Eq}[1]{Eq.~(\ref{#1})}
\newcommand{\Eqs}[2]{Eqs.~(\ref{#1})--(\ref{#2})}

\begin{document}

\title{Multipair-resilient entanglement swapping with complementary
linear-optical measurements}

\author{Siavash Mirzaei Ghormish}
\affiliation{Department of Electrical and Computer Engineering, Brigham Young University, Provo, Utah 84602, USA}

\author{Ryan M. Camacho}
\affiliation{Department of Electrical and Computer Engineering, Brigham Young University, Provo, Utah 84602, USA}

\date{\today}

\begin{abstract}
Multipair emission is a principal source of false heralds in entanglement
swapping with spontaneous parametric down-conversion sources. A conventional
Bell-state measurement (BSM) cannot distinguish the desired arrival of one
photon from each neighboring source from a mixed-polarization double emission by
one source accompanied by vacuum from the other. We show that a cascaded network
can resolve this ambiguity by allowing its successive swapping stations to
perform different measurements. A direct-basis BSM rejects same-polarization
double emissions, whereas a balanced equatorial analyzer, implemented by a
four-mode Green Machine (GM), uses two-photon interference to reject
mixed-polarization double emissions. At the same time, it recovers
same-polarization inter-source events rejected by the BSM as resolved
$\phi$-type Bell heralds. Their rejection sets jointly cover both
single-source two-photon classes while retaining contributions from both useful
inter-source classes.
We classify the passive four-mode, number-resolving analyzers satisfying this
condition and identify the BSM and GM as canonical balanced representatives.
Using a Gaussian-state analysis exact to all orders of multipair emission, we
evaluate three-source BSM--GM and four-source BSM--GM--BSM Pure Bell Pair sources
under coupling, detector, and channel loss. The three-source network suppresses
the leading infidelity, while the four-source network heralds an exact Bell state
in the lossless limit, a property that holds for every alternating BSM--GM chain
with four or more sources. The fidelity advantage over source-count-matched
all-BSM networks persists at every loss level studied, a single-channel rate
advantage appears without multiplexing, and spectral multiplexing raises the
Bell-pair delivery probability toward the asymptotically deterministic limit.
\end{abstract}

\maketitle

\section{Introduction}
\label{sec:intro}

Quantum repeaters extend entanglement by joining elementary links through
entanglement swapping~\cite{kimble2008quantum,wehner2018quantum,
lloyd2004infrastructure,azuma2023repeaters,sangouard2011quantum}. Their performance
depends on how often a Bell-state measurement (BSM) announces success and on
whether each herald really marks a single pair emitted by each neighboring
source. Errors in
the pairs those links deliver compound even when subsequent swaps are ideal. For example,
$N$ identical Werner links of singlet fidelity $F$ produce an end-to-end fidelity
$F_N=[1+3((4F-1)/3)^N]/4$~\cite{dur1999quantum}; achieving $F_N=0.9$ across ten
links requires approximately $F=0.99$ per link. Purification can restore fidelity,
but only by consuming additional pairs and storing them during repeated rounds,
while memory lifetime is already a bottleneck in current repeaters
~\cite{bennett1996purification,dur1999quantum,ustc2026}. High-rate operation
depends on each swap heralding a high-fidelity pair directly.

The standard SPDC-based photonic swap does not meet that requirement. A
linear-optical BSM heralds on an orthogonally polarized two-photon coincidence,
its interference erasing which-source information. But the same detector record
can be produced by a mixed-polarization double emission from one source while
the neighboring source emits vacuum. The BSM cannot authenticate source history. For thermal
SPDC pair statistics, this ambiguity caps both the fidelity of the heralded
state and the probability that a herald marks a true Bell pair at $1/2$ in the
ideal lossless limit. Memory-assisted
postselection can remove the vacuum component, but a fidelity of $0.99$ then
requires a mean pair number per mode below ${\sim}0.01$, so successful
generation is rare~\cite{sangouard2011quantum,shapiro2026entanglement}. Multiplexing recovers rate by
pumping many spectral, temporal, or spatial modes in parallel
~\cite{sinclair2014spectral,chakraborty2025towards,chen2023zalm,
shapiro2024entanglement,shapiro2025high}, but it leaves the reliability of each
individual herald unchanged. The same ambiguity arises for any source with a
nonzero multipair probability, including four-wave-mixing and quantum-dot sources
~\cite{garay2007photon,arakawa2020progress,liu2019solid,
bassobasset2019entanglement,zopf2019entanglement}.

The problem is easiest to state in the two-photon input space. Label the two rails
entering a swapping station $a$ and $b$, one arriving from each neighboring
source, and the polarizations $H,V$. Within the two-photon manifold, the ten
normalized Fock states fall into four orthogonal symmetry classes:
\begin{equation}
\begin{array}{c|cc}
 &\text{different pol.}&\text{same pol.}\\ \hline
\text{different rails}&\mathcal C_1\ (\text{desired})&\mathcal C_3\\
\text{same rail}&\mathcal C_2&\mathcal C_4.
\end{array}
\label{eq:intro_classes}
\end{equation}
Representative states are $a_H^\dagger b_V^\dagger\ket0$ in $\mathcal C_1$,
$a_H^\dagger a_V^\dagger\ket0$ in $\mathcal C_2$,
$a_H^\dagger b_H^\dagger\ket0$ in $\mathcal C_3$, and
$(a_H^\dagger)^2\ket0$ in $\mathcal C_4$. The conventional BSM target lies in
$\mathcal C_1$, but both different-rail classes $\mathcal C_1$ and
$\mathcal C_3$ contain one photon from each source. The two same-rail classes $\mathcal C_2$ and
$\mathcal C_4$ are the signatures of a double emission from a single source and
must not survive the complete measurement network. A BSM does not resolve
$\mathcal C_3$; a complementary analyzer, introduced below, converts that
otherwise discarded valid sector into resolved $\phi$-type
($\ket{\Phi^\pm}$-like) Bell heralds.

Extending entanglement across a chain of three sources requires two swapping
measurements, one at each of the two intermediate stations. These two
operations, which we label $A$ and $B$, are conventionally implemented as
identical Bell-state measurements, but nothing requires this. Chahine et al. showed that
alternating direct- and diagonal-basis BSMs suppresses multipair false heralds
since a double-pair emission cannot independently trigger both adjacent
measurements, and that the alternating protocol permits all four Bell
outcomes~\cite{chahine2023protocol}. We ask a broader question: which
analyzer pairs provide optimal rejection? Let $P_k^{(A)}$ and
$P_k^{(B)}$ denote the acceptance probabilities
of operations $A$ and $B$ for a state in class $\mathcal C_k$, with $P_k=0$
meaning that every state in the class is rejected and $P_k>0$ that some state is
accepted. The pair must satisfy
\begin{equation}
\begin{gathered}
P_1^{(A)}P_1^{(B)}>0,\\
P_k^{(A)}P_k^{(B)}=0\quad(k=2,4).
\end{gathered}
\label{eq:intro_covering}
\end{equation}
Each single-source class must be rejected completely by at least one operation,
while some desired amplitude must pass both.

Solving this covering problem from first principles yields three main results.
First, all entangling solutions belong to one structural family: a direct-basis
polarization-parity analyzer paired with an equatorial-basis analyzer. Balanced
mixing with maximally entangled heralds selects the conventional BSM and Guha's four-mode Hadamard
joint-detection receiver: a Green Machine (GM)~\cite{guha2011structured};
Chahine \emph{et al.}'s protocol is its diagonal special case. Second, every
lossless alternating BSM--GM chain with four or more sources heralds an exact
Bell state to all SPDC emission orders, without receiver postselection. Third,
an all-orders Gaussian analysis includes detector, coupling, and channel loss
and quantifies the resulting fidelity and rate advantages.
The BSM measures direct-basis ($H/V$) polarization parity (it accepts exactly one $H$ and
one $V$ detection): it retains
$\mathcal C_1$ but also accepts the same-rail mixed-polarization class $\mathcal C_2$,
while rejecting both same-polarization classes $\mathcal C_3$ and $\mathcal C_4$.
The GM performs the complementary operation. Its diagonal-basis polarization
analysis and two-photon interference retain part of $\mathcal C_1$ and reject
$\mathcal C_2$ exactly, while $\mathcal C_3$ and $\mathcal C_4$ can pass. Their
rejection sets therefore cover both single-source classes:
\begin{equation}
{\setlength{\arraycolsep}{9pt}
\begin{array}{c|cccc}
 &\mathcal C_1&\mathcal C_2&\mathcal C_3&\mathcal C_4\\ \hline
\mathrm{BSM}&>0&>0&0&0\\
\mathrm{GM}&>0&0&>0&>0.
\end{array}}
\label{eq:intro_filter_map}
\end{equation}
Here $>0$ denotes a nonzero accepted-sector probability, whereas $0$ denotes
exact rejection. For a polarization-entangled source, each term of a double
emission presents the same class, $\mathcal C_2$ or $\mathcal C_4$, to the
analyzers on its two sides (Appendix~\ref{appendix:fock_state_analysis}), so at
least one of them rejects it. The GM's acceptance of $\mathcal C_3$ requires one
photon from each neighbor and heralds a $\phi$-type Bell state. The GM thus
removes the BSM's $\mathcal C_2$ false positives and recovers its
$\mathcal C_3$ false negatives.

This classification holds among lossless passive four-mode analyzers with
photon-number-resolving detection and a two-bank acceptance rule, after
excluding measurements that preserve which-source information and therefore
cannot swap. Appendix~\ref{app:analyzer_classification} gives the proof; below
we describe the physical action of its balanced BSM--GM representatives.

We call the resulting complementary architecture a Pure Bell Pair (\PBP)
source. We analyze a three-source BSM--GM cascade and its symmetric four-source
BSM--GM--BSM extension, benchmarking each against a source-count-matched all-BSM
cascade under the same detector, coupling, and channel loss. The three-source
network removes the vacuum-dominated false-herald background of the all-BSM
chain and leaves a single higher-order residual; the four-source network removes
that residual and heralds an exact Bell state in the lossless limit, a property
we show holds, to all orders of emission, for every alternating BSM--GM chain
with four or more sources. Under the matched comparison, the PBP network maintains a
fidelity advantage at every loss level studied and a single-channel rate
advantage without multiplexing; multiplexing magnifies the rate advantage,
making the three-source transmitter near-deterministic and, in the lossless
limit, the four-source transmitter asymptotically deterministic in the sense
defined in Sec.~\ref{sec:fom}.

\section{Complementary measurements and PBP architecture}
\label{sec:arch}

\subsection{The BSM--GM measurement pair}

\begin{figure}[tbp!]
    \centering
    \begin{tikzpicture}[
    x=0.95cm,
    y=0.72cm,
    optical/.style={draw, line width=0.75pt},
    hwp/.style={draw, fill=cyan!35, minimum width=0.48cm,
        minimum height=0.09cm, inner sep=0pt},
    note/.style={font=\scriptsize},
    paneltitle/.style={font=\small\bfseries, anchor=west}
]

\newcommand{\simpledetector}[1]{%
    \begin{scope}[shift={#1}]
        \draw[line width=0.75pt]
            (0,-0.24) -- (0,0.24)
            arc[start angle=90,end angle=-90,radius=0.24] -- cycle;
    \end{scope}%
}

\newcommand{\simpleanalyzer}[2]{%
    \node[paneltitle] at (0,4.45) {#1};
    \node[note, anchor=east] at (0.55,3.15) {$a$};
    \node[note, anchor=east] at (0.55,1.15) {$b$};
    \draw[optical] (0.65,3.15) -- (1.80,3.15);
    \draw[optical] (0.65,1.15) -- (1.80,1.15);

    \coordinate (bs) at (2.65,2.15);
    \coordinate (pbsu) at (4.05,3.28);
    \coordinate (pbsl) at (4.05,1.02);
    \draw[optical] (1.80,3.15) -- (bs) -- (pbsl);
    \draw[optical] (1.80,1.15) -- (bs) -- (pbsu);
    \draw[panelbg, line width=3.2pt] (2.30,2.15) -- (3.00,2.15);
    \fill[black] (2.30,2.10) rectangle (3.00,2.20);
    \node[note, anchor=east] at (2.22,2.15) {$50{:}50$ BS};

    #2

    \draw[optical] (pbsu) -- (4.68,3.82) -- (5.85,3.82);
    \draw[optical] (pbsu) -- (4.68,2.74) -- (5.85,2.74);
    \draw[optical] (pbsl) -- (4.68,1.48) -- (5.85,1.48);
    \draw[optical] (pbsl) -- (4.68,0.40) -- (5.85,0.40);
    \draw[panelbg, line width=3.2pt] (3.70,3.28) -- (4.40,3.28);
    \fill[black] (3.70,3.23) rectangle (4.40,3.33);
    \draw[panelbg, line width=3.2pt] (3.70,1.02) -- (4.40,1.02);
    \fill[black] (3.70,0.97) rectangle (4.40,1.07);
    \node[note, anchor=east] at (3.62,3.55) {PBS};
    \node[note, anchor=east] at (3.62,0.75) {PBS};

    \simpledetector{(5.85,3.82)}
    \simpledetector{(5.85,2.74)}
    \simpledetector{(5.85,1.48)}
    \simpledetector{(5.85,0.40)}
}

\begin{scope}[shift={(0,0)}]
\colorlet{panelbg}{white}
\filldraw[fill=panelbg, draw=black, line width=0.75pt]
    (-0.20,0.02) rectangle (6.78,4.78);
\simpleanalyzer{(a) BSM ($H/V$ basis)}{}
\node[note, anchor=west] at (6.12,3.82) {$c_H$};
\node[note, anchor=west] at (6.12,2.74) {$c_V$};
\node[note, anchor=west] at (6.12,1.48) {$d_H$};
\node[note, anchor=west] at (6.12,0.40) {$d_V$};
\end{scope}

\begin{scope}[shift={(0,-5.15)}]
\colorlet{panelbg}{cyan!12}
\filldraw[fill=panelbg, draw=black, line width=0.75pt]
    (-0.20,0.02) rectangle (6.78,4.78);
\simpleanalyzer{(b) GM (equatorial basis)}{
    \node[hwp, rotate=-58.5] at (3.40,2.75) {};
    \node[hwp, rotate=58.5] at (3.40,1.55) {};
    \node[note, anchor=west] at (3.52,2.48) {$22.5^\circ$ HWP};
    \node[note, anchor=west] at (3.52,1.82) {$22.5^\circ$ HWP};
}
\node[note, anchor=west] at (6.12,3.82) {$c_+$};
\node[note, anchor=west] at (6.12,2.74) {$c_-$};
\node[note, anchor=west] at (6.12,1.48) {$d_+$};
\node[note, anchor=west] at (6.12,0.40) {$d_-$};
\end{scope}

\end{tikzpicture}
    \caption{Analyzer layouts. Both mix the input rails $a$ and $b$ on a 50:50
    beam splitter (BS). The BSM resolves $H/V$ directly at a polarizing beam
    splitter (PBS) in each output rail; the GM inserts a $22.5^\circ$ half-wave
    plate (HWP) before each PBS and resolves $+/-$. Their complementary
    decisions are summarized in Table~\ref{tab:complementary_cases}.}
    \label{fig:analyzer_schematics}
\end{figure}
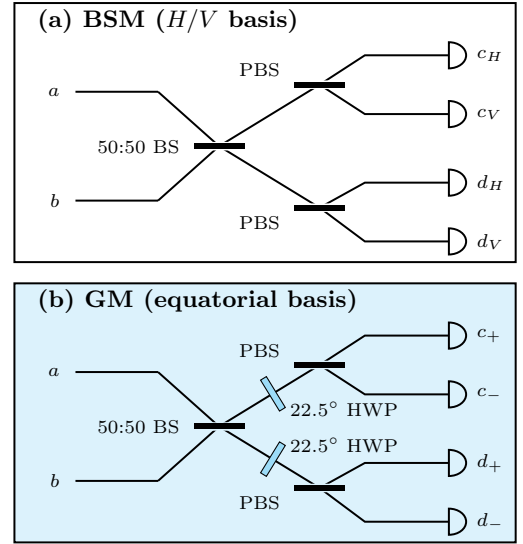

Figure~\ref{fig:analyzer_schematics} shows the two analyzer layouts. Although
their hardware is nearly identical, they accept different two-photon inputs. A
conventional BSM first mixes the input rails on a 50:50 beam splitter and then
resolves $H/V$ polarization in each output rail. It accepts the records that
contain one $H$ and one $V$ detection. This rejects the same-polarization classes
$\mathcal C_3$ and $\mathcal C_4$ (for $\mathcal C_3$, this is the familiar
restriction of a linear-optical BSM to the $\psi^\pm$ sector), but it cannot
determine whether an accepted
$H/V$ pair entered through different rails ($\mathcal C_1$) or through the same
rail ($\mathcal C_2$).

The GM uses the same balanced rail mixing but analyzes both output rails in the
equatorial $+/-$ basis, implemented by a half-wave plate at $22.5^\circ$ followed
by a PBS.  The two indistinguishable amplitudes that lead to an accepted $+/-$
record interfere, so the decision depends on the joint two-photon state. For the
same-rail $H/V$ input in $\mathcal C_2$, those amplitudes cancel
for every accepted record; for the different-rail $H/V$ input in $\mathcal C_1$,
some accepted amplitudes survive. The GM resolves the BSM's
source-origin ambiguity in the mixed-polarization sector. Its remaining blind spot
is the same-rail same-polarization class $\mathcal C_4$, which the BSM already
rejects. Its acceptance of $\mathcal C_3$ is equally important: a $\mathcal C_3$
input carries one photon from each source, and the GM's same-rail records herald
it as a $\phi$-type Bell state (Sec.~\ref{sec:three}). For the balanced analyzers,
the valid-sector probabilities are
\[
\bigl(P_1^{\rm BSM},P_3^{\rm BSM}\bigr)=(1,0),
\qquad
\bigl(P_1^{\rm GM},P_3^{\rm GM}\bigr)=\left(\tfrac12,\tfrac12\right).
\]
The GM therefore leaves the total acceptance of the equal-weight inter-source
sector unchanged and redistributes it: half of the $\mathcal C_1$ events are
exchanged for $\mathcal C_3$ events that the BSM would discard.

Pairwise source correlations make this recovery constructive at the network
level. In the three-source chain, two coherent histories can present
$(b_1,a_2)=(V,H)$ and $(b_2,a_3)=(V,V)$, or instead $(H,V)$ and $(H,H)$.
The BSM sees $\mathcal C_1$ in both histories while the neighboring GM sees
$\mathcal C_3$; the unmeasured endpoints are respectively $HH$ and $VV$.
Because the two histories are coherent, erasing which one occurred leaves the
endpoints in the pure superposition $\ket{\phi^\pm}$.

For either analyzer, we order the detector bank from top to bottom and accept the
four photon-number records
\begin{equation}
\mathcal A=\{1001,0110,1100,0011\}.
\label{eq:accepted_records}
\end{equation}
For the BSM the ordering is $(c_H,c_V,d_H,d_V)$; for the GM it is
$(c_+,c_-,d_+,d_-)$. Table~\ref{tab:complementary_cases} enumerates the complete
two-photon input basis and shows the complementary blind spots of the two
analyzers.

\begin{table*}[t!]
    \centering
    {%
\small
\setlength{\tabcolsep}{4pt}
\renewcommand{\arraystretch}{1.18}
\setlength{\fboxsep}{3pt}
\newcommand{\rh}[1]{\multicolumn{1}{c}{$[#1]$}}
\begin{tabular}{@{}c l cccc cc cccc c@{}}
\toprule
& & \multicolumn{4}{c}{\textbf{Accepted set} $\mathcal A$} &
\multicolumn{2}{c}{\textbf{Same-outcome}} &
\multicolumn{4}{c}{\textbf{Doubles} $\mathcal D$} & \\
\cmidrule(lr){3-6}\cmidrule(lr){7-8}\cmidrule(lr){9-12}
\textbf{Class} & \textbf{Two-photon input}
 & \rh{1100} & \rh{0011} & \rh{1001} & \rh{0110}
 & \rh{1010} & \rh{0101}
 & \rh{2000} & \rh{0200} & \rh{0020} & \rh{0002}
 & \textbf{Decision ($P_{\rm acc}$)} \\
\midrule
\multicolumn{13}{@{}l}{\fbox{\textbf{BSM}}\;\; records ordered $(c_H,c_V,d_H,d_V)$}\\
\midrule
\multirow{2}{*}{$\mathcal C_1$}
 & $a_H^\dagger b_V^\dagger\ket0$ & $\tfrac14$ & $\tfrac14$ & $\tfrac14$ & $\tfrac14$ & & & & & & & $\psi$-type herald $(1)$ \\
 & $a_V^\dagger b_H^\dagger\ket0$ & $\tfrac14$ & $\tfrac14$ & $\tfrac14$ & $\tfrac14$ & & & & & & & $\psi$-type herald $(1)$ \\
\midrule
\multirow{2}{*}{$\mathcal C_2$}
 & $a_H^\dagger a_V^\dagger\ket0$ & $\tfrac14$ & $\tfrac14$ & $\tfrac14$ & $\tfrac14$ & & & & & & & false herald $(1)$ \\
 & $b_H^\dagger b_V^\dagger\ket0$ & $\tfrac14$ & $\tfrac14$ & $\tfrac14$ & $\tfrac14$ & & & & & & & false herald $(1)$ \\
\midrule
\multirow{2}{*}{$\mathcal C_3$}
 & $a_H^\dagger b_H^\dagger\ket0$ & & & & & & & $\tfrac12$ & & $\tfrac12$ & & rejected $(0)$ \\
 & $a_V^\dagger b_V^\dagger\ket0$ & & & & & & & & $\tfrac12$ & & $\tfrac12$ & rejected $(0)$ \\
\midrule
\multirow{4}{*}{$\mathcal C_4$}
 & $(a_H^\dagger)^2\ket0$ & & & & & $\tfrac12$ & & $\tfrac14$ & & $\tfrac14$ & & rejected $(0)$ \\
 & $(a_V^\dagger)^2\ket0$ & & & & & & $\tfrac12$ & & $\tfrac14$ & & $\tfrac14$ & rejected $(0)$ \\
 & $(b_H^\dagger)^2\ket0$ & & & & & $\tfrac12$ & & $\tfrac14$ & & $\tfrac14$ & & rejected $(0)$ \\
 & $(b_V^\dagger)^2\ket0$ & & & & & & $\tfrac12$ & & $\tfrac14$ & & $\tfrac14$ & rejected $(0)$ \\
\midrule
\multicolumn{13}{@{}l}{\fcolorbox{black}{cyan!12}{\textbf{GM}}\;\; records ordered $(c_+,c_-,d_+,d_-)$}\\
\midrule
\multirow{2}{*}{$\mathcal C_1$}
 & $a_H^\dagger b_V^\dagger\ket0$ & & & $\tfrac14$ & $\tfrac14$ & & & $\tfrac18$ & $\tfrac18$ & $\tfrac18$ & $\tfrac18$ & $\psi$-type herald $(1/2)$ \\
 & $a_V^\dagger b_H^\dagger\ket0$ & & & $\tfrac14$ & $\tfrac14$ & & & $\tfrac18$ & $\tfrac18$ & $\tfrac18$ & $\tfrac18$ & $\psi$-type herald $(1/2)$ \\
\midrule
\multirow{2}{*}{$\mathcal C_2$}
 & $a_H^\dagger a_V^\dagger\ket0$ & & & & & $\tfrac14$ & $\tfrac14$ & $\tfrac18$ & $\tfrac18$ & $\tfrac18$ & $\tfrac18$ & rejected $(0)$ \\
 & $b_H^\dagger b_V^\dagger\ket0$ & & & & & $\tfrac14$ & $\tfrac14$ & $\tfrac18$ & $\tfrac18$ & $\tfrac18$ & $\tfrac18$ & rejected $(0)$ \\
\midrule
\multirow{2}{*}{$\mathcal C_3$}
 & $a_H^\dagger b_H^\dagger\ket0$ & $\tfrac14$ & $\tfrac14$ & & & & & $\tfrac18$ & $\tfrac18$ & $\tfrac18$ & $\tfrac18$ & $\phi$-type herald $(1/2)$ \\
 & $a_V^\dagger b_V^\dagger\ket0$ & $\tfrac14$ & $\tfrac14$ & & & & & $\tfrac18$ & $\tfrac18$ & $\tfrac18$ & $\tfrac18$ & $\phi$-type herald $(1/2)$ \\
\midrule
\multirow{4}{*}{$\mathcal C_4$}
 & $(a_H^\dagger)^2\ket0$ & $\tfrac18$ & $\tfrac18$ & $\tfrac18$ & $\tfrac18$ & $\tfrac18$ & $\tfrac18$ & $\tfrac1{16}$ & $\tfrac1{16}$ & $\tfrac1{16}$ & $\tfrac1{16}$ & false herald $(1/2)$ \\
 & $(a_V^\dagger)^2\ket0$ & $\tfrac18$ & $\tfrac18$ & $\tfrac18$ & $\tfrac18$ & $\tfrac18$ & $\tfrac18$ & $\tfrac1{16}$ & $\tfrac1{16}$ & $\tfrac1{16}$ & $\tfrac1{16}$ & false herald $(1/2)$ \\
 & $(b_H^\dagger)^2\ket0$ & $\tfrac18$ & $\tfrac18$ & $\tfrac18$ & $\tfrac18$ & $\tfrac18$ & $\tfrac18$ & $\tfrac1{16}$ & $\tfrac1{16}$ & $\tfrac1{16}$ & $\tfrac1{16}$ & false herald $(1/2)$ \\
 & $(b_V^\dagger)^2\ket0$ & $\tfrac18$ & $\tfrac18$ & $\tfrac18$ & $\tfrac18$ & $\tfrac18$ & $\tfrac18$ & $\tfrac1{16}$ & $\tfrac1{16}$ & $\tfrac1{16}$ & $\tfrac1{16}$ & false herald $(1/2)$ \\
\bottomrule
\end{tabular}
}

    \caption{Complete two-photon case classification, organized like
    Eq.~\eqref{eq:intro_filter_map}. Rows enumerate the ten creation-operator
    inputs, grouped into the four classes of Eq.~\eqref{eq:intro_classes};
    columns list all ten possible two-photon detector records, grouped into the
    accepted set $\mathcal A$ of Eq.~\eqref{eq:accepted_records}, the two
    same-outcome cross-rail records, and the four double records
    $\mathcal D$. Entries give the probability of each record for the given
    input (blank entries are zero); the final column gives the decision and the
    total acceptance probability $P_{\rm acc}$ summed over $\mathcal A$. The
    complementary structure is visible directly: the BSM's $\mathcal C_1$ and
    $\mathcal C_2$ rows are identical (its source-origin ambiguity), while every
    $\mathcal C_2$ input leaves the GM's $\mathcal A$ columns empty (exact
    rejection). The complementary
    rejection pattern covers both single-source classes, $\mathcal C_2$ (rejected
    by the GM) and $\mathcal C_4$ (rejected by the BSM). Inputs in $\mathcal C_3$
    carry one photon from each rail; the BSM does not resolve them, whereas the
    GM's same-rail records herald them as $\phi$-type Bell states
    (Sec.~\ref{sec:three}). Thus the balanced GM trades half of the BSM's
    $\mathcal C_1$ acceptance for half of the otherwise discarded $\mathcal C_3$
    sector while rejecting the $\mathcal C_2$ false herald exactly.}
    \label{tab:complementary_cases}
\end{table*}

\subsection{The complementary analyzer family}
\label{sec:general}

The alternating-BSM protocol suppresses multipair contributions by pairing
direct- and rotated-basis measurements~\cite{chahine2023protocol}. In a
different source architecture, Marcellino \emph{et al.} cascade single-photon
path-entangled states and use a rotated central BSM to suppress false heralds
from two orthogonally polarized photons arriving from the same
side~\cite{joseph2024toward}. The class-covering condition developed here is
more general: it characterizes the broader family of analyzer pairs whose
complementary rejection patterns jointly eliminate both single-source
two-photon classes while retaining the useful inter-source sectors.
Appendix~\ref{app:analyzer_classification} shows that, within the architecture
of Sec.~\ref{sec:intro} (a passive four-mode unitary, PNR detection, and one
photon in each of two detector banks), every entangling solution pairs a
direct-basis polarization-parity analyzer with an analyzer that measures an
equatorial polarization basis in each input rail; the equatorial phases may
differ between the rails, and the detector modes within each bank may be rotated
freely. These freedoms leave the rejection zeros of
Eq.~\eqref{eq:intro_filter_map} unchanged. Requiring in addition that every
accepted record project onto a maximally entangled state selects balanced rail
mixing, which gives the conventional BSM and, for the equatorial analyzer, the
GM. The rest of this subsection writes the balanced family explicitly.

For the balanced, rail-symmetric subfamily used throughout the performance
analysis, the direct polarization basis
$\mathcal B_Z=\{\ket H,\ket V\}$ is mutually unbiased with every basis in the
equatorial family~\cite{wootters1989optimal,durt2010mutually}
\begin{equation}
    \ket{d_{\phi,\pm}}=\frac{\ket H\pm e^{-i\phi}\ket V}{\sqrt2},
    \qquad 0\le\phi<2\pi,
    \label{eq:MUB_equatorial_basis}
\end{equation}
because each equatorial state has squared overlap $1/2$ with both $H$ and $V$.
The transformation that maps $H/V$ onto this basis is
\begin{equation}
\Rphi=\frac{1}{\sqrt2}\begin{pmatrix}1 & e^{i\phi}\\[2pt] 1 & -e^{i\phi}\end{pmatrix},
\qquad 0\le\phi<2\pi.
\label{eq:Rphi_intro}
\end{equation}
Combining it with balanced rail mixing gives the four-mode analyzer
\begin{equation}
\Uphi=B_{\rm rail}\otimes\Rphi,\qquad
B_{\rm rail}=\frac{1}{\sqrt2}\begin{pmatrix}1&1\\1&-1\end{pmatrix},
\label{eq:Uphi_def}
\end{equation}
where $B_{\rm rail}$ is the balanced two-rail mixer. Because
$\Rphi=R_0\,\mathrm{diag}(1,e^{i\phi})$ and $B_{\rm rail}=R_0=H_2$, one has
\begin{equation}
\Uphi=U_0\,[\,I_{\rm rail}\otimes\mathrm{diag}(1,e^{i\phi})\,],\qquad
U_0=H_2\otimes H_2 ,
\label{eq:Uphi_gm}
\end{equation}
i.e., the GM is the real-valued member $U_0=H_2\otimes H_2$, while $\phi$ is an
input phase on the vertical modes. Every input mode has equal-modulus overlap
with every detector mode, but the tensor-product phase structure fixes the
two-photon interference that performs the filtering.

Writing a four-mode input as
$\ket{n_{aH},n_{aV};n_{bH},n_{bV}}$, every $\Uphi$ rejects the same-rail
mixed-polarization inputs $\ket{1,1;0,0}$ and $\ket{0,0;1,1}$ from
$\mathcal C_2$, while same-polarization inputs such as $\ket{2,0;0,0}$ and
$\ket{0,0;2,0}$ can pass. The BSM has the opposite rejection pattern, giving the
class coverage in Eq.~\eqref{eq:intro_filter_map}. The factorization in
Eq.~\eqref{eq:Uphi_gm} shows that varying the common phase $\phi$ changes neither
the destructive-interference zeros nor their probabilities. More generally, the
independent rail phases allowed by Eq.~\eqref{eq:classification_physical_family}
give the desired Bell component
\begin{equation}
\ket{\psi^-}\longrightarrow
\frac{\ket{HV}-e^{i(\phi_a-\phi_b)}\ket{VH}}{\sqrt2},
\label{eq:phase_rotated_bell}
\end{equation}
up to an overall phase. The phase difference is removed by a known local
$Z(\phi_a-\phi_b)$ correction. In the rail-symmetric family used in the
performance calculations, $\phi_a=\phi_b=\phi$, so the desired Bell state itself
is unchanged; only phases relative to orthogonal multiphoton sectors vary. This
permits the analyzer to absorb a stable common birefringent phase without changing
fidelity, success probability, or loss behavior. We use the canonical $\phi=0$
GM throughout.

\subsection{Source and measurement network}

We use the complementary pair as the swap primitive of a polarization-entangled
quantum transmitter (QTX). Each Sagnac-configured SPDC source emits
polarization-entangled photon pairs in the singlet state, and neighboring source
outputs are joined by alternating BSM and GM stages. The
source pulses are synchronized to a common clock. A cascade herald is
accepted only when every stage registers an allowed record in the same spectral
channel and pump pulse, so the network performs one global entanglement swap
without intermediate memories.

\begin{figure*}[t]
    \centering
    \begin{tikzpicture}[
    optical/.style={draw, line width=0.75pt},
    inputpath/.style={optical, -{Latex[length=1.7mm]}},
    source/.style={draw, line width=0.75pt, fill=blue!8,
        minimum width=1.42cm, minimum height=1.12cm,
        align=center, font=\scriptsize},
    analyzer/.style={draw, line width=0.75pt, fill=white,
        minimum width=1.55cm, minimum height=0.88cm,
        align=center, font=\scriptsize\bfseries},
    gm/.style={analyzer, fill=cyan!12},
    rail/.style={font=\scriptsize, fill=white, inner sep=1.2pt},
    note/.style={font=\scriptsize, align=center},
    endpoint/.style={font=\scriptsize\bfseries}
]

\newcommand{\downphotodetector}[1]{%
    \begin{scope}[shift={#1}, rotate=-90]
        \draw[line width=0.7pt]
            (0,-0.17) -- (0,0.17)
            arc[start angle=90,end angle=-90,radius=0.17] -- cycle;
    \end{scope}%
}

\coordinate (s1) at (1.55,5.10);
\coordinate (m1) at (3.75,5.10);
\coordinate (s2) at (5.95,5.10);
\coordinate (mg) at (8.15,5.10);
\coordinate (s3) at (10.35,5.10);
\coordinate (m2) at (12.55,5.10);
\coordinate (s4) at (14.75,5.10);

\node[source] at (s1) {Sagnac 1\\$\ket{\psi^-}$};
\node[source] at (s2) {Sagnac 2\\$\ket{\psi^-}$};
\node[source] at (s3) {Sagnac 3\\$\ket{\psi^-}$};
\node[source] at (s4) {Sagnac 4\\$\ket{\psi^-}$};

\node[analyzer] (bsmone) at (m1) {$\mathrm{BSM}_1$\\$H/V$};
\node[gm]       (gman)   at (mg) {$\mathrm{GM}$\\$+/-$};
\node[analyzer] (bsmtwo) at (m2) {$\mathrm{BSM}_2$\\$H/V$};

\draw[optical, -{Latex[length=1.7mm]}] (0.84,5.10) -- (0.15,5.10);
\draw[optical, -{Latex[length=1.7mm]}] (15.46,5.10) -- (16.15,5.10);
\node[endpoint, anchor=east] at (0.08,5.10) {Alice};
\node[endpoint, anchor=west] at (16.22,5.10) {Bob};
\node[rail] at (0.50,5.10) {$a_1$};
\node[rail] at (15.82,5.10) {$b_4$};

\draw[inputpath] (2.26,5.10) -- (bsmone.west);
\draw[inputpath] (5.24,5.10) -- (bsmone.east);
\draw[inputpath] (6.66,5.10) -- (gman.west);
\draw[inputpath] (9.64,5.10) -- (gman.east);
\draw[inputpath] (11.06,5.10) -- (bsmtwo.west);
\draw[inputpath] (14.04,5.10) -- (bsmtwo.east);

\node[rail] at (2.70,5.36) {$b_1$};
\node[rail] at (4.78,5.36) {$a_2$};
\node[rail] at (7.10,5.36) {$b_2$};
\node[rail] at (9.18,5.36) {$a_3$};
\node[rail] at (11.50,5.36) {$b_3$};
\node[rail] at (13.58,5.36) {$a_4$};

\foreach \stage/\mx in {bsmone/3.75,gman/8.15,bsmtwo/12.55}{
    \foreach \dx in {-0.60,-0.20,0.20,0.60}{
        \draw[optical]
            ($({\stage}.south)+(\dx,0)$) -- ({\mx+\dx},4.19);
        \downphotodetector{({\mx+\dx},4.19)}
    }
    \node[note] at (\mx,3.50) {PNR detectors};
}

\end{tikzpicture}
    \caption{Four-source \PBP\ transmitter. Two outer BSMs connect source pairs
    1--2 and 3--4, while the central GM connects sources 2--3. Each black line is
    one spatial rail containing both polarization modes; the surviving outer rails
    $a_1$ and $b_4$ are delivered to Alice and Bob. The four resolved outputs of
    each analyzer are monitored by photon-number-resolving detectors. Removing
    Sagnac~4 and $\mathrm{BSM}_2$ gives the three-source BSM--GM transmitter,
    whose right endpoint is $b_3$.}
    \label{fig:arch}
\end{figure*}
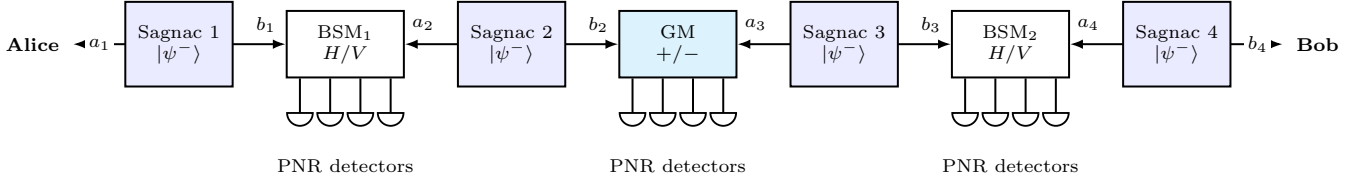

Figure~\ref{fig:arch} shows the full four-source BSM--GM--BSM network.

The $i$th source has signal rail $a_i$ and idler rail $b_i$, each with $H$ and
$V$ polarization modes. In the three-source transmitter, $b_1$ and $a_2$ enter
$\mathrm{BSM}_1$, while $b_2$ and $a_3$ enter the GM; $a_1$ and $b_3$ remain as
the delivered modes. In the four-source transmitter, a second BSM joins $b_3$ and
$a_4$, and the surviving modes are $a_1$ and $b_4$. We use the BSM--GM--BSM
ordering throughout; interchanging the two analyzer types gives the complementary
GM--BSM--GM ordering.

The probabilities computed below are joint per-pulse quantities evaluated on the
global optical state: because there are no intermediate memories, every stage
must succeed in the same pulse. This removes storage and its decoherence penalty,
but it makes the raw per-pulse success probability small. Spectral multiplexing
recovers that probability. We assume pulse
synchronization and path-length matching sufficient for photons from different
sources to reach each analyzer within their mutual coherence time.

\subsection{Spectral multiplexing and heralding}

Each Sagnac source is engineered to emit $N_I$ spectrally separated,
phase-matched islands. Pump-envelope shaping and
$\chi^{(2)}$-crystal engineering provide established routes to approximately
factorable islands~\cite{Uren2005,Mosley2008,Branczyk2011,Fejer1992,Hum2007}, each
of which is modeled as an independent two-mode squeezed-vacuum source. The
measurement outputs are demultiplexed by dense wavelength-division multiplexing
(DWDM) into these $N_I$ channels. Spectral multiplexing raises the aggregate
heralding probability while allowing the brightness, and hence the multipair
probability, to remain low in each channel.

All detectors are assumed to resolve photon number, so the accepted patterns below
are photon-number records. Threshold detectors do not implement the analyzed
heralding rule: higher-order emissions can
produce an accepted threshold pattern while leaving the outer modes in vacuum.

At each stage, the four records in Eq.~\eqref{eq:accepted_records} comprise a
\emph{split} family $S=\{1001,0110\}$, with one photon in each output rail, and a
\emph{same-rail} family $B=\{1100,0011\}$, with one photon in each polarization
detector of a single output rail. A same-channel cascade herald occurs only when
every stage records one of these patterns in the same DWDM channel $n$ and pump
pulse. The conditioned outer state is then associated with that resolved channel.
Because the channels are approximately orthogonal, the device acts as $N_I$
parallel heralded generators.

An accepted detector record is necessary but not sufficient to herald a Bell pair,
because multipair emission can reproduce the same record;
Secs.~\ref{sec:three} and~\ref{sec:four} quantify the resulting false-herald
background of the matched \allBSM\ chains and its suppression by complementary
filtering.

\subsection{Figures of merit and comparison model}
\label{sec:fom}

\begin{table*}[t]
\centering
\begin{tabular*}{\textwidth}{@{\extracolsep{\fill}}lll@{}}
\toprule
\textbf{Symbol} & \textbf{Meaning} & \textbf{Defined} \\
\midrule
$P_{\rm gen}$ & per-channel, per-pattern generation probability & \Eq{eq:Fph} \\
$P_{\rm Herald}=4^{j-1}P_{\rm gen}$ & per-channel herald probability & \Eqs{eq:Pherald3}{eq:Pherald4} \\
$B_{\rm Ph}$ & Bell-subspace fraction of a herald & \Eq{eq:Bph} \\
$\calF_{\rm Ph}=P_{\psi^-}/P_{\rm gen}$ & photon--photon fidelity & \Eq{eq:Fph} \\
$P_{\rm Success}$ & multiplexed Bell-subspace success probability & \Eq{eq:Psuccess} \\
$B_S$ & Bell-subspace fraction after loading & \Eq{eq:Fs} \\
$\calF_{S}$ & spin--spin fidelity after loading & \Eq{eq:Fs} \\
$\RE$;\ $P_{\rm ent}=\RE/\RP$ & delivery rate; per-pulse entanglement probability & \Eqs{eq:RE_general}{eq:Pent} \\
\bottomrule
\end{tabular*}
\caption{Figures of merit. Probability entries are defined per pump pulse;
$P_{\rm gen}$ and $P_{\rm Herald}$ additionally refer to one spectral channel.
The index $j\in\{3,4\}$ is the source count.}
\label{tab:figures_of_merit}
\end{table*}

We use two transmitter-level metrics for the heralded flying-photon state and two
receiver-level metrics conditioned on successful memory loading
(Table~\ref{tab:figures_of_merit}). Photon--photon
fidelity and multiplexed success probability characterize the transmitter; spin--spin
fidelity and entanglement-delivery rate characterize the loaded state at Alice and
Bob. Throughout, each \PBP\ network is compared with the \allBSM\ chain having
the same source count, brightness, multiplexing, detector model, and loss model
(hereafter the matched \allBSM\ chain). Only the measurement network changes.

The photon--photon fidelity is the fraction of heralded events carrying the target
singlet $\ket{\psi^-}$ after the record-dependent local correction,
\begin{equation}
    \calF_{\rm Ph}^{(j)} = \frac{P_{\psi^-}^{(j)}}{P_{\rm gen}^{(j)}},
    \qquad j\in\{3,4\},
    \label{eq:Fph}
\end{equation}
where $j$ is the number of Sagnac sources, $P_{\psi^-}^{(j)}$ is the joint
per-channel, per-pattern, per-pulse probability of a herald and the target state,
and $P_{\rm gen}^{(j)}$ is the corresponding probability of that herald regardless
of the outer state. The Bell fraction
\begin{equation}
    B_{\rm Ph}^{(j)}=\frac{P_{\rm Bell}^{(j)}}{P_{\rm gen}^{(j)}}
    \label{eq:Bph}
\end{equation}
counts any state in the two-photon Bell subspace, whether or not it is the
target singlet. The distinction matters because a known local correction can convert
between heralded Bell states, whereas vacuum and same-side two-photon events are
not usable Bell pairs. Under internal loss, $B_{\rm Ph}$ also counts
one-photon-per-side events whose polarization state has been mixed by the loss of
a companion photon (Sec.~\ref{sec:three}), so it is an upper bound on the usable
Bell fraction; the lossless success probabilities below are unaffected.

Each stage has four accepted records and, by symmetry, every combination of
records is equiprobable, so the per-channel heralding probabilities of the three-
and four-source transmitters are
\begin{equation}
    P_{\rm Herald}^{(3)} = 4^2\,P_{\rm gen}^{(3)} = 16\,P_{\rm gen}^{(3)},
    \label{eq:Pherald3}
\end{equation}
\begin{equation}
    P_{\rm Herald}^{(4)} = 4^3\,P_{\rm gen}^{(4)} = 64\,P_{\rm gen}^{(4)}.
    \label{eq:Pherald4}
\end{equation}
For every GM stage, these factors include both split records, which herald
$\psi$-type states, and both same-rail records, which herald $\phi$-type states.
All are retained using the resolved record and the corresponding local Pauli
correction.
With same-channel multiplexing, there are $N_I$ statistically identical trials per
pump pulse. We select an available channel using its herald record, without
measuring the outer photons. The probability that this selected herald carries a
state in the Bell-pair subspace is
\begin{equation}
    P_{\rm Success}^{(j)} =
    B_{\rm Ph}^{(j)}\left[1-\left(1-P_{\rm Herald}^{(j)}\right)^{N_I}\right].
    \label{eq:Psuccess}
\end{equation}

At the quantum receiver (QRX), successful loading postselects events with at least one photon
in Alice's memory and at least one in Bob's. Conditional on a transmitter herald, let
$P_{\rm Loadable}^{(j)}$ be the probability of successful loading and
$P_{\rm Bell,load}^{(j)}$ and $P_{\psi^-,{\rm load}}^{(j)}$ the probabilities of
successful loading in the Bell subspace and in the target singlet, respectively.
The receiver Bell fraction and spin--spin fidelity are
\begin{equation}
    B_S^{(j)}=\frac{P_{\rm Bell,load}^{(j)}}{P_{\rm Loadable}^{(j)}},\qquad
    \calF_{S}^{(j)} = \frac{P_{\psi^-,{\rm load}}^{(j)}}{P_{\rm Loadable}^{(j)}}.
    \label{eq:Fs}
\end{equation}
Let $K_j$ be the number of successful
same-channel heralds among the $N_I$ channels; since the channels are independent,
\begin{equation}
    \Pr(K_j) = \binom{N_I}{K_j}\left(P_{\rm Herald}^{(j)}\right)^{K_j}
    \left(1-P_{\rm Herald}^{(j)}\right)^{N_I-K_j}.
    \label{eq:binomial}
\end{equation}
If $N_M$ memories are available per pump pulse, only $\min(K_j,N_M)$ heralds are
usable, so the mean number of usable heralds is
\begin{equation}
    \overline{K}_j(N_M) = \sum_{K_j=0}^{N_I}\min(K_j,N_M)\,\Pr(K_j),
    \qquad N_M\le N_I,
    \label{eq:Kbar}
\end{equation}
and the same-channel entanglement-delivery rate is
\begin{equation}
    \RE^{(j)}(N_M) = \RP\,\overline{K}_j(N_M)\,P_{\psi^-,{\rm load}}^{(j)},
    \qquad j\in\{3,4\},
    \label{eq:RE_general}
\end{equation}
with $\RP$ the pump repetition rate; $P_{\psi^-,{\rm load}}^{(j)}$ depends on
the propagation and receiver efficiency $\etaR$ through
Appendix~\ref{app:three}. For single-memory operation ($N_M=1$),
\begin{equation}
    \RE^{(j)}(1) = \RP\left[1-\left(1-P_{\rm Herald}^{(j)}\right)^{N_I}\right]
    P_{\psi^-,{\rm load}}^{(j)},
    \label{eq:RE_single}
\end{equation}
and when memories match channels ($N_M=N_I$), every herald is usable and
\begin{equation}
    \RE^{(j)}(N_I) = \RP\,N_I\,P_{\rm Herald}^{(j)}\,P_{\psi^-,{\rm load}}^{(j)}.
    \label{eq:RE_full}
\end{equation}
Finally, the per-pulse entanglement probability is
\begin{equation}
    P_{\rm ent} = \frac{\RE}{\RP}.
    \label{eq:Pent}
\end{equation}
For single-memory operation, we adopt the following operational terminology: a
source is \emph{quasi-deterministic} when $P_{\rm ent}$ reaches the conventional
BSM-limited value $1/2$, and \emph{near-deterministic} when
$P_{\rm ent}\ge0.75$. We call the source \emph{asymptotically deterministic}
when $P_{\rm ent}\to1$ as the multiplexing increases. In the lossless case with
$N_M=1$, $P_{\rm ent}$ coincides with $P_{\rm Success}$, so the same terms apply
to the success probabilities of Figs.~\ref{fig:Psucc3} and~\ref{fig:Psucc4}.

\section{Three-source PBP source}
\label{sec:three}

The three-source transmitter is the minimal cascade in which the two intermediate
stations can perform different measurements: a BSM joins sources 1 and 2, and a
GM joins sources 2 and 3. A two-source swap has only one measurement station and
therefore cannot apply both complementary filters; replacing its BSM with a GM
only exchanges one blind spot for the other. The matched BSM--BSM cascade is the
appropriate benchmark. We first obtain the conditioned
lossless state, then evaluate fidelity and usable rate under loss. The derivations are given in
Appendices~\ref{app:three} and~\ref{appendix:fock_state_analysis}.

\subsection{Lossless heralded state}

For any accepted BSM record, a split GM record $S$ conditions the outer modes
$a_1$ and $b_3$ into the normalized state
\begin{equation}
\begin{aligned}
\ket{\psi_{S}}
={}&\frac{1}{\sqrt{2(2P_1+P_2)}}\\[-2pt]
&\times\Big[\sqrt{2P_{1}}
 \left(\ket{1,0;0,1}\pm\ket{0,1;1,0}\right)\\
&\qquad\mathrel{\pm}\sqrt{P_{2}}
 \left(\ket{1,1;0,2}-\ket{1,1;2,0}\right)\Big]_{a_1b_3},
\end{aligned}
\label{eq:psiS3}
\end{equation}

whereas a same-rail GM record $B$ gives
\begin{equation}
\begin{aligned}
\ket{\psi_{B}}
={}&\frac{1}{\sqrt{2(2P_1+P_2)}}\\[-2pt]
&\times\Big[\sqrt{2P_{1}}
 \left(\ket{1,0;1,0}\pm\ket{0,1;0,1}\right)\\
&\qquad\mathrel{\pm}\sqrt{P_{2}}
 \left(\ket{1,1;0,2}-\ket{1,1;2,0}\right)\Big]_{a_1b_3}.
\end{aligned}
\label{eq:psiB3}
\end{equation}

Here $P_{n}=(G-1)^{n}/G^{n+1}$ is the Bose--Einstein probability of $n$ pairs per
mode. The gain is $G$, so $G-1$ is the mean pair number per mode per pulse. Our
Fock-state convention includes the normalization
$\ket{2}=(a^\dagger)^2\ket{0}/\sqrt2$. Because every accepted record contains
exactly two photons, photon-number conservation restricts lossless accepted
events to the pair-number histories $(q,2-q,q)$ with $q\in\{0,1,2\}$
(Appendix~\ref{appendix:fock_state_analysis}); Eqs.~\eqref{eq:psiS3}
and~\eqref{eq:psiB3} are therefore exact to all orders of emission. The
per-record generation probability, the squared
norm of the unnormalized heralded state derived in
Appendix~\ref{appendix:fock_state_analysis}, is given in
Eq.~\eqref{eq:Pgen3_lossless}.

A split record heralds the $\ket{\psi^\pm}$ subspace and a same-rail record the
$\ket{\phi^\pm}$ subspace; the resolved BSM and GM records determine the signs
and hence the required local Pauli correction. The two record families are
equiprobable in the balanced network, so both contribute equally to the heralding
and delivery rates. In both cases the first term is the
desired Bell contribution and the second is a residual higher-order contribution.
Their relative weights give
\begin{align}
\calF_{\rm Ph}^{\rm PBP}
&=\frac{2P_1}{2P_1+P_2}
 =\frac{2G}{3G-1},\nonumber\\
1-\calF_{\rm Ph}^{\rm PBP}
&=\frac{G-1}{3G-1}.
\label{eq:Fph3_lossless}
\end{align}
The photon--photon infidelity is approximately $(G-1)/2$ at low brightness.
For example, at $G-1=0.005$ the desired Bell weight exceeds the residual weight
by $2P_1/P_2\approx402$.

For comparison, with $s_i\in\{+1,-1\}$ fixed by the resolved records, the
matched BSM--BSM chain heralds
\begin{equation}
\begin{aligned}
\ket{\psi_{\rm BSM}}
={}&\frac{1}{\sqrt{2(P_0+2P_1+P_2)}}\\[-2pt]
&\times\Big[\sqrt{2P_{1}}
 \left(\ket{1,0;0,1}+s_1\ket{0,1;1,0}\right)\\
&\qquad+s_2\sqrt{2P_0}\ket{0,0;0,0}\\
&\qquad+s_3\sqrt{2P_2}\ket{1,1;1,1}\Big]_{a_1b_3}.
\end{aligned}
\label{eq:psiBSM3}
\end{equation}
The additional vacuum term dominates as $G-1\rightarrow0$, so the
pre-postselection Bell fraction is
\begin{equation}
B_{\rm Ph}^{\rm BSM}
=\frac{2P_1}{P_0+2P_1+P_2}
\approx 2(G-1),
\label{eq:Bph3_BSM_lossless}
\end{equation}
whereas $B_{\rm Ph}^{\rm PBP}=2P_1/(2P_1+P_2)\rightarrow1$. Ideal memory loading
removes the all-BSM vacuum term, after which both architectures have fidelity
$2P_1/(2P_1+P_2)$. The difference is where that fidelity is obtained: the PBP
network produces it directly in the flying state, whereas the all-BSM network
reaches it only after receiver postselection has discarded most of its heralds. Their surviving multipair
terms also differ, which leads to different loss dependence below.

The contrast between Eqs.~\eqref{eq:psiS3} and~\eqref{eq:psiBSM3} is the
network-level consequence of Sec.~\ref{sec:arch}: the GM removes the BSM's
mixed-polarization false-herald class, so the PBP state has no vacuum term. The
remaining contribution survives only because sources 1 and 3 are endpoints: a
mixed-polarization double pair from source 1 passes $\mathrm{BSM}_1$ through
$\mathcal C_2$, and a same-polarization double pair from source 3 passes the GM
through $\mathcal C_4$, each accompanied by vacuum from source 2. It places two
photons on each side, one more than the Bell term, and is smaller than the Bell
amplitude by $\mathcal O(\sqrt{P_2/P_1})$. It can be flagged by local
photon-number resolution and is eliminated by the four-source construction of
Sec.~\ref{sec:four}.

\subsection{Fidelity under loss}

The Gaussian-state calculation in Appendix~\ref{app:three} includes multipair
emission to all orders. Detector and heralding-path losses are combined into the
effective internal transmissions $\etaT$ and $\etaG$ for the BSM and GM,
respectively. The receiver transmission $\etaR$ accounts for propagation and
coupling of the outer modes into Alice's and Bob's memories.

\begin{figure}[tbp]
    \centering
    \includegraphics{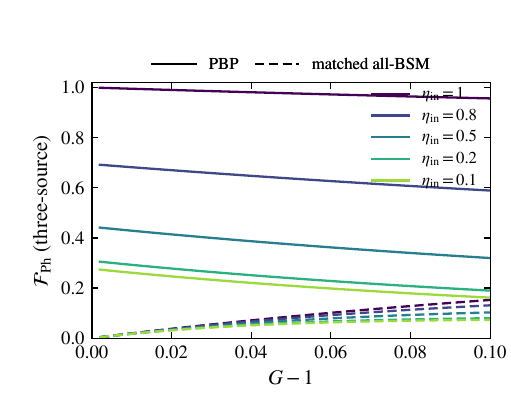}
    \caption{Photon--photon fidelity $\calF_{\rm Ph}$ versus mean photon number
    $G-1$ for internal efficiencies $\etaT=\etaG=\eta_{\rm in}=1, 0.8, 0.5, 0.2, 0.1$,
    three-source. Solid: \PBP\ network (BSM--GM). Dashed: matched
    \allBSM\ chain (BSM--BSM). The \PBP\ fidelity approaches unity at low
    brightness for all efficiencies, whereas the \allBSM\ fidelity collapses toward
    zero because its pre-postselection state is vacuum-dominated.}
    \label{figure2}
\end{figure}

Figure~\ref{figure2} shows the photon--photon fidelity for equal internal
transmissions, which we denote $\eta_{\rm in}=\etaT=\etaG$ throughout. The PBP
fidelity approaches unity at
low brightness and remains well above the all-BSM value throughout the plotted
range. Internal loss lowers both fidelities at fixed brightness because higher
emissions can lose photons and imitate an accepted two-photon record. The
advantage produced by complementary filtering persists at every plotted
transmission.

\begin{figure}[tbp]
    \centering
    \includegraphics{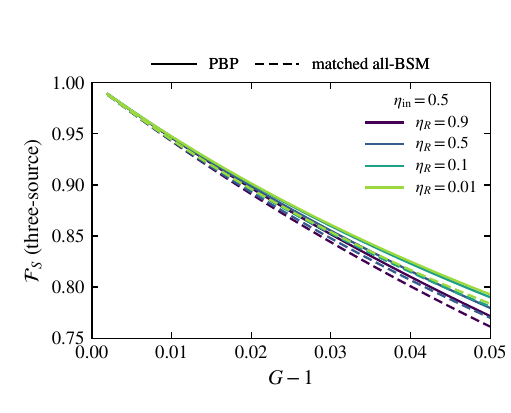}
    \caption{Spin--spin fidelity $\calF_S$ versus $G-1$ for output transmissions
    $\etaR=0.9,0.5,0.1,0.01$, three-source, $\eta_{\rm in}=0.5$. Solid: \PBP; dashed:
    matched \allBSM\ chain. The \PBP\ network lies above the matched chain
    throughout; the modest rise with decreasing $\etaR$ arises from the
    reweighting of histories admitted by internal loss (see text).}
    \label{figure3}
\end{figure}

\begin{figure}[tbp]
    \centering
    \includegraphics{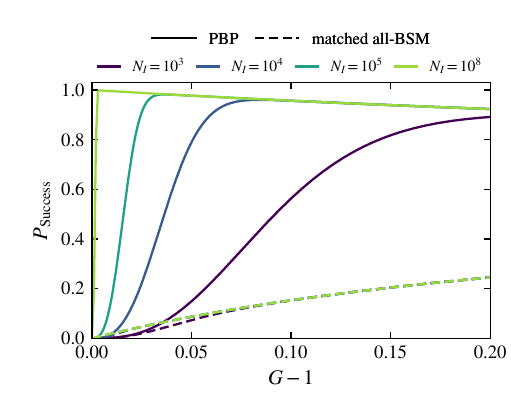}
    \caption{Success probability $P_{\rm Success}$ versus $G-1$ for spectral
    multiplexing $N_I=10^{3},10^{4},10^{5},10^{8}$ (bottom to top), three-source,
    lossless. Solid: \PBP; dashed: matched \allBSM\
    chain. The \PBP\ network reaches near-deterministic operation at high multiplexing;
    the \allBSM\ chain remains below the quasi-deterministic threshold.}
    \label{fig:Psucc3}
\end{figure}

Figure~\ref{figure3} gives the spin--spin fidelity after successful memory loading
for $\eta_{\rm in}=0.5$. The roles of receiver loss and internal loss can be
separated analytically. Set $\eta_{\rm in}=1$, so that
Eq.~\eqref{eq:psiS3} is the complete heralded state, and write
$q=1-\etaR$. Its Bell component loads with probability $\etaR^2$, whereas its
two-photon-per-side component loads with probability
$[1-q^2]^2=\etaR^2(2-\etaR)^2$. Exactly one photon survives on each side with
probability $[2\etaR q]^2$. The lost photons retain the polarization information
that distinguishes these alternatives, so the surviving two-qubit state is
maximally mixed: all of this weight lies in the Bell subspace, but only one quarter
overlaps the target singlet. It follows that
\begin{align}
B_S(\etaR)
&=\frac{2P_1+4P_2(1-\etaR)^2}
        {2P_1+P_2(2-\etaR)^2},\nonumber\\
\calF_S(\etaR)
&=\frac{2P_1+P_2(1-\etaR)^2}
        {2P_1+P_2(2-\etaR)^2}.
\label{eq:receiver_loss_ideal3}
\end{align}
Thus $B_S\rightarrow1$ as $\etaR\rightarrow0^+$, but the fidelity approaches
$(2P_1+P_2)/(2P_1+4P_2)$, slightly below its $\etaR=1$ value
$2P_1/(2P_1+P_2)$. Receiver loss alone therefore does not purify the ideal
heralded state.

The modest rise in conditional fidelity with decreasing $\etaR$ in
Fig.~\ref{figure3} has a different origin. Internal loss admits additional
higher-emission histories into the heralded ensemble, and receiver loss changes
their relative weight in the loadable sector. This conditional reweighting comes
at the expense of rate: the probability that both memories load still falls with
$\etaR$. Across the plotted transmissions, the PBP source retains less non-Bell
weight in the loadable sector and sustains the higher fidelity as brightness
increases. At the $\eta_{\rm in}=0.5$ of Fig.~\ref{figure3} and
$\etaR=0.01$, for example, the operating points at $\calF_S=0.99$ are
$G-1=1.80\times10^{-3}$ for PBP and $1.69\times10^{-3}$ for the all-BSM chain.

\subsection{Success probability and delivery rate}

The lossless generation probabilities also clarify the rate tradeoff. For one
resolved record, the squared norms of the unnormalized heralded states
(Appendix~\ref{appendix:fock_state_analysis}) are
\begin{align}
P_{\rm gen}^{\rm PBP}
&=\frac{P_0^3P_1^2(2P_1+P_2)}{16},\nonumber\\
P_{\rm gen}^{\rm BSM}
&=\frac{P_0^3P_1^2(P_0+2P_1+P_2)}{16}.
\label{eq:Pgen3_lossless}
\end{align}
Multiplying each by its Bell fraction gives the same single-channel true-Bell
probability. This equality has a direct class-level interpretation: on the valid
different-rail sector the BSM accepts all of $\mathcal C_1$ and none of
$\mathcal C_3$, whereas the balanced GM accepts half of each. Pair correlations
give the two classes equal total weight, so the recovered $\mathcal C_3$
$\phi$-type heralds exactly compensate the reduced $\mathcal C_1$ acceptance.
In the lossless, unmultiplexed limit, complementary filtering therefore preserves
the true-event probability while removing the much larger false-herald background
of the all-BSM chain. With multiplexing the difference appears: as
$N_I\rightarrow\infty$, Eq.~\eqref{eq:Psuccess} approaches the Bell-fraction
ceiling, which tends to unity for PBP but only to $2(G-1)$ for the all-BSM chain.
At $G-1=0.005$, those ceilings differ by approximately a factor of 100.

Loss lowers the Bell-subspace event probability for both networks and increases
the number of channels required to approach either ceiling. Even without
multiplexing, complementary rejection improves the usable single-channel yield
under loss. At the fidelity-matched operating
points used below ($\eta_{\rm in}=\etaR=0.7$ and $\calF_S=0.99$), the $N_I=1$
delivery probability is approximately $1.12$ times larger for PBP; for
$\eta_{\rm in}=\etaR=0.7$, $N_I=10^7$, and $G-1=0.005$, the PBP success
probability is $0.366$, compared with $0.00975$ for the all-BSM chain.

Because all stages must herald in the same channel and pump pulse, synchronous
operation has a small per-channel probability. The multiplexed probability begins
to approach its ceiling when
\begin{equation}
N_I P_{\rm Herald}^{(j)}\gtrsim1.
\label{eq:NI_requirement}
\end{equation}
Internal loss and each additional swap stage reduce $P_{\rm Herald}^{(j)}$ and
so increase the required mode count. This cost comes from eliminating
intermediate memories and applies equally to the matched all-BSM chain, which is
subject to the same synchronous condition. Purely spectral
$N_I\sim10^7$ is technologically demanding, so a practical implementation would
likely combine time, frequency, and spatial modes; the formulas apply with $N_I$
replaced by the total number of independent same-pulse trials.

\begin{figure}[tbp]
    \centering
    \includegraphics{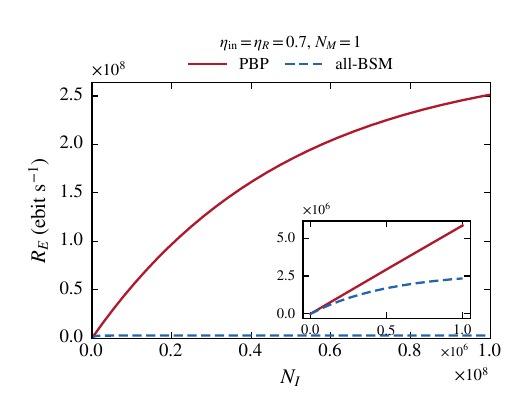}
    \caption{Entanglement-delivery rate $\RE$ (ebit/s) versus spectral-island
    count $N_I$, three-source, single-memory ($N_M=1$). Solid: \PBP; dashed:
    matched \allBSM\ chain. Parameters: $\RP=10^{9}\,\mathrm{s}^{-1}$,
    $\eta_{\rm in}=\etaR=0.7$; $G-1=2.95\times10^{-3}$ (\PBP) and
    $2.84\times10^{-3}$ (\allBSM). Inset: low-multiplexing detail.}
    \label{fig:RE3single}
\end{figure}

\begin{figure}[tbp]
    \centering
    \includegraphics{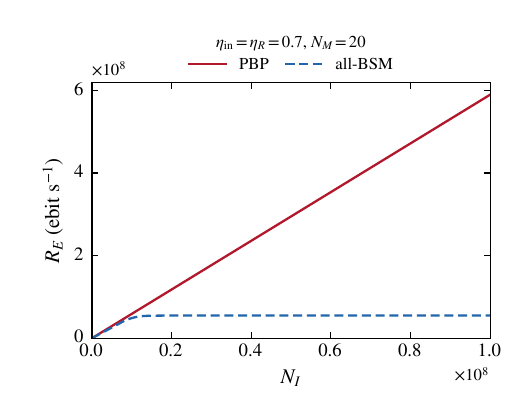}
    \caption{Entanglement-delivery rate $\RE$ (ebit/s) versus $N_I$, three-source,
    multi-memory ($N_M=20$). Solid: \PBP; dashed: matched \allBSM\
    chain. Parameters as in Fig.~\ref{fig:RE3single}. Additional memories raise both
    rates and push saturation to higher $N_I$; the \PBP\ network saturates well beyond
    the \allBSM\ chain.}
    \label{fig:RE3multi}
\end{figure}

Figures~\ref{fig:RE3single} and~\ref{fig:RE3multi} give the entanglement-delivery
rate for $N_M=1$ and $N_M=20$ receiver memories. We impose
$\calF_S=0.99$ and take $\eta_{\rm in}=\etaR=0.7$ and
$\RP=10^{9}\,\mathrm{s}^{-1}$. The PBP and all-BSM operating points are
$G-1=2.95\times10^{-3}$ and $2.84\times10^{-3}$, respectively. With one memory,
the all-BSM rate saturates near $N_I\sim10^6$ at $2.7\times10^6$~ebit/s, whereas
the PBP rate continues to rise beyond the plotted range and saturates only near
$N_I\sim2\times10^8$. At $N_I=10^7$ and $10^8$, the PBP rates are
$5.3\times10^7$ and $2.5\times10^8$~ebit/s, approximately 19 and 90 times the
all-BSM values. With $N_M=20$, the corresponding PBP rates are
$5.9\times10^7$ and $5.9\times10^8$~ebit/s, approximately 1.2 and 11 times the
all-BSM values.

The single-memory curves show where the advantage comes from. The all-BSM herald
probability is dominated by vacuum false heralds, so the bracket in
Eq.~\eqref{eq:RE_single} saturates already near $N_I\sim10^6$; but the one
herald per pulse that a single memory can accept is then a vacuum event with
probability $\approx1-B_{\rm Ph}$, and the rate saturates at
$\RP P_{\psi^-,{\rm load}}\approx2.7\times10^6$~ebit/s. The PBP network
heralds about two orders of magnitude less often and needs correspondingly more
channels to saturate, but nearly every herald it selects carries a Bell pair, so
its saturated rate is roughly two orders of magnitude higher. With $N_M=20$ the
all-BSM chain can
store several heralds per pulse and becomes memory-limited at
$N_M\RP P_{\psi^-,{\rm load}}$, which is why the ratio falls to 1.2--11; in the
lossless, fully multiplexed limit $N_M=N_I$, Eqs.~\eqref{eq:RE_full}
and~\eqref{eq:Pgen3_lossless} give equal rates for the two networks.

The transmissions used here correspond to a representative link length as
follows. Writing $\eta_{\rm in}=\eta_{\rm d}10^{-\alpha L/10}$ for each
source-to-station segment and $\etaR=\eta_{\rm c}10^{-\alpha L'/10}$ for each
output arm, detector efficiency $\eta_{\rm d}=0.9$, combined coupling and
conversion efficiency $\eta_{\rm c}=0.9$, and fiber attenuation
$\alpha=0.2$~dB/km give $L=5.5$~km per internal segment and $L'=5.5$~km per
output arm. The three-source geometry spans an internal section of
$4L=21.8$~km and two output arms totaling $2L'=10.9$~km, for a total span of
approximately 33~km.

\section{Four-source PBP source}
\label{sec:four}

The four-source transmitter (Fig.~\ref{fig:arch}) combines four Sagnac sources
with the symmetric BSM--GM--BSM network. In the derivation, the two outer BSMs are
applied first, joining sources 1--2 and 3--4 and leaving the intermediate modes
$b_2$ and $a_3$ to be fused by the central GM. This construction addresses the
only higher-order term left by the
three-source cascade: each BSM-prepared joint system contains a same-side
two-photon component, and the central GM applies the complementary filter to the
pair of joint systems. Appendix~\ref{appendix:fock_state_analysis} gives the
Fock-state derivation.

\subsection{Lossless heralded state}

For a split GM record $S$, the conditioned outer modes $a_1$ and $b_4$ are in the
exact Bell state
\begin{equation}
\ket{\psi_{S}}
=\frac{1}{\sqrt2}\left(\ket{1,0;0,1}-\ket{0,1;1,0}\right)_{a_1b_4},
\label{eq:psiS4}
\end{equation}
with per-record generation probability $P_{\rm gen}^{(4)}=P_0^4P_1^4/32$, the
squared norm of the unnormalized state derived in
Appendix~\ref{appendix:fock_state_analysis}.

Split GM records herald $\ket{\psi^{\pm}}$, whereas same-rail records herald
$\ket{\phi^{\pm}}$; the two BSM records fix the sign (Table~\ref{tab:outer}).
All four Bell states are therefore accessible and can be converted to a chosen
target by a record-dependent local correction; a relative equatorial phase
between the rails would add only the known local phase of
Eq.~\eqref{eq:phase_rotated_bell}.

\begin{table}[tb]
    \centering
    \footnotesize
    \renewcommand{\arraystretch}{1.25}
    \setlength{\tabcolsep}{4pt}
    \begin{tabular}{c|cccc|cccc}
        \hline
             & \multicolumn{4}{c|}{GM $=S$} & \multicolumn{4}{c}{GM $=B$}\\
        \hline
        BSM$_1$ & $S$ & $S$ & $B$ & $B$ & $S$ & $S$ & $B$ & $B$\\
        BSM$_2$ & $S$ & $B$ & $S$ & $B$ & $S$ & $B$ & $S$ & $B$\\
        \hline
        Outer & $\ket{\psi^-}$ & $\ket{\psi^+}$ & $\ket{\psi^+}$ & $\ket{\psi^-}$
             & $\ket{\phi^-}$ & $\ket{\phi^+}$ & $\ket{\phi^+}$ & $\ket{\phi^-}$\\
        \hline
    \end{tabular}
    \caption{Heralded outer Bell state versus record combination. $S$ denotes
    the split records $\{1001,0110\}$ and $B$ the same-rail records
    $\{1100,0011\}$; the rail-symmetric Green Machine gives the entries as shown.
    The $S$ and $B$ blocks have equal total probability, and both are included in
    the reported rates after a record-dependent local Pauli correction.}
    \label{tab:outer}
\end{table}

The cancellation can be read directly from the symmetry classes. Each
BSM-prepared joint system carries a contribution
$\propto\ket{1,1;0,0}\pm\ket{0,0;1,1}$ on its outer and inner rails
($a_1b_2$ for sources 1--2 and $a_3b_4$ for sources
3--4)~\cite{dhara2022heralded}. At the central station the inner-rail part is
the mixed-polarization, same-input class $\mathcal C_2$, which the GM rejects.
The symmetric cascade therefore removes the residual term present in
Eqs.~\eqref{eq:psiS3} and~\eqref{eq:psiB3}.

The same argument extends to any alternating chain. Because each accepted record
contains exactly two photons, photon-number conservation restricts lossless
accepted events to the pair-number histories $(q,2-q,q,\ldots)$ with
$q\in\{0,1,2\}$. The $q=1$ history is the desired one; the $q=0$ and $q=2$
histories each contain a double-pair source lying between a BSM and a GM, and the
three two-pair terms of such a source are each annihilated by one analyzer or the
other [Eq.~\eqref{eq:sagnac_two_pair_classes}]. Hence every alternating BSM--GM
chain with $N\geq4$ sources heralds a pure Bell state to all emission orders under
lossless, number-resolving operation
(Appendix~\ref{appendix:fock_state_analysis}; Table~\ref{tab:finite_chain_check}).
For $N=3$ the double-pair sources are endpoints, which is why the residual in
Eqs.~\eqref{eq:psiS3} and~\eqref{eq:psiB3} survives.

For comparison, the matched \allBSM\ chain gives
\begin{equation}
\begin{split}
\ket{\psi_{S}} =\; & \frac12
\Bigg[\left(\ket{1,0;0,1}_{a_{1}b_{4}}\pm\ket{0,1;1,0}_{a_{1}b_{4}}\right)\\
&\pm\left(\ket{1,1;0,0}_{a_{1}b_{4}}-\ket{0,0;1,1}_{a_{1}b_{4}}\right)\Bigg],
\end{split}
\label{eq:psiBSM4}
\end{equation}
with per-record generation probability $P_0^4P_1^4/16$, twice that of the PBP
network. The second parenthesis contains both outer photons on the same side and
contributes no usable flying Bell pair. Its weight equals that of the
desired first parenthesis, giving the lossless all-BSM chain
$\calF_{\rm Ph}=B_{\rm Ph}=1/2$; the doubled herald rate consists entirely of
false heralds. The four-source PBP state has unit fidelity. Accordingly, in the
lossless high-multiplexing limit its success
probability approaches unity, whereas the matched all-BSM chain approaches
$1/2$ and requires receiver postselection to discard its same-side events.

\subsection{Fidelity and architecture tradeoff}
\begin{figure}[tbp]
    \centering
    \includegraphics{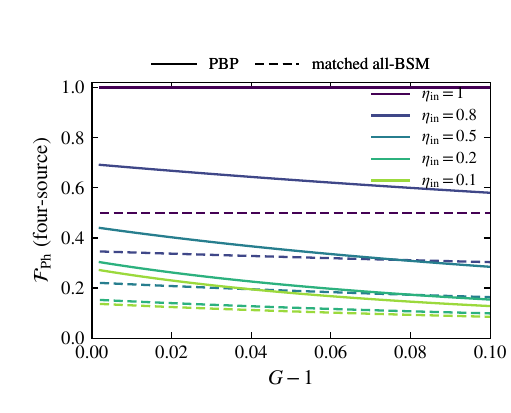}
    \caption{Photon--photon fidelity $\calF_{\rm Ph}$ versus $G-1$ for
    $\etaT=\etaG=\eta_{\rm in}=1, 0.8, 0.5, 0.2, 0.1$, four-source. Solid: \PBP\
    (BSM--GM--BSM); dashed: matched \allBSM\ chain (BSM--BSM--BSM). The
    \PBP\ fidelity reaches unity in the ideal limit; the \allBSM\ fidelity is capped
    at $0.5$.}
    \label{fig:Fph4}
\end{figure}

The Gaussian-state calculation in Appendix~\ref{app:four} includes emission to
all orders and loss in all three measurement stages. Figure~\ref{fig:Fph4} shows
the photon--photon fidelity when those stages have a common effective
transmission $\eta_{\rm in}$. In the ideal low-brightness limit, PBP approaches
unit fidelity
while the matched all-BSM chain approaches its $1/2$ ceiling. Increasing
brightness or internal loss admits additional multipair histories into the
accepted detector records and lowers both fidelities. Because the four-source
network contains one more source and measurement stage than the three-source
network, it pays a correspondingly larger loss penalty.

\begin{figure}[tbp]
    \centering
    \includegraphics{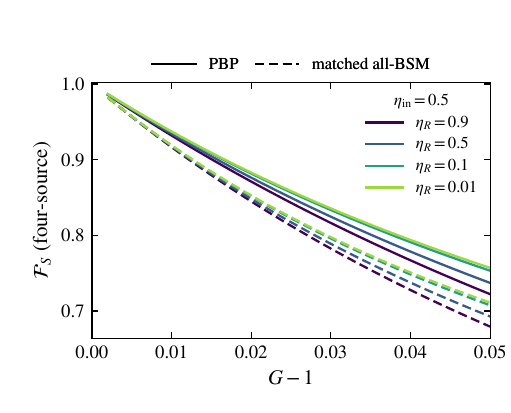}
    \caption{Spin--spin fidelity $\calF_S$ versus $G-1$ for
    $\etaR=0.9,0.5,0.1,0.01$, four-source, $\eta_{\rm in}=0.5$. Solid: \PBP; dashed:
    matched \allBSM\ chain. As in Fig.~\ref{figure3}, the \PBP\ network lies
    above the matched chain and the conditional fidelity rises modestly as
    $\etaR$ decreases.}
    \label{fig:Fs4}
\end{figure}

Figure~\ref{fig:Fs4} shows the postselected
spin--spin fidelity versus $G-1$ for several $\etaR$ at $\eta_{\rm in}=0.5$; as in
the three-source case, lower $\etaR$ raises this conditional fidelity and the \PBP\
network stays above the \allBSM\ chain. As Eq.~\eqref{eq:receiver_loss_ideal3}
shows, receiver loss does not purify the heralded state; the ideal four-source
state is an exact singlet, whose $\calF_S$ it cannot change at all. The rise
arises in the internally lossy ensemble, where receiver loss reweights the
additional higher-emission histories admitted by internal loss. At the fidelity
target $\calF_S=0.99$ with $\etaR=\eta_{\rm in}=0.7$, PBP can operate at
$G-1=2.71\times10^{-3}$, compared with $2.16\times10^{-3}$ for the all-BSM
chain.

\begin{figure}[tbp]
    \centering
    \includegraphics{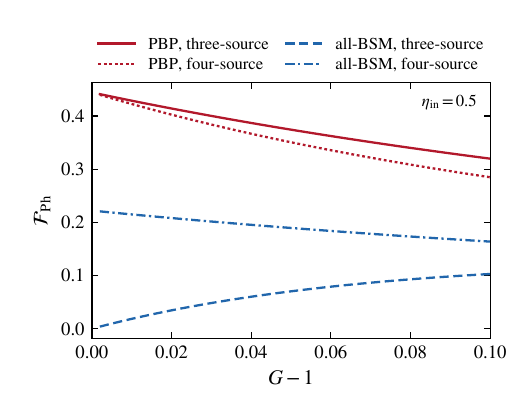}
    \caption{Photon--photon fidelity $\calF_{\rm Ph}$ versus $G-1$ at
    $\etaT=\etaG=\eta_{\rm in}=0.5$, comparing three- and four-source
    networks (\PBP\ and \allBSM). The three-source \PBP\ network lies above its
    four-source counterpart; the ordering is reversed for the \allBSM\ chains
    because the three-source chain carries the vacuum term of
    Eq.~\eqref{eq:psiBSM3}.}
    \label{fig:Fph34}
\end{figure}

\begin{figure}[tbp]
    \centering
    \includegraphics{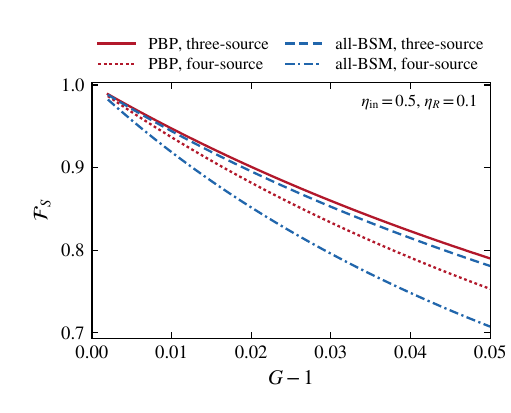}
    \caption{Spin--spin fidelity $\calF_S$ versus $G-1$ at $\eta_{\rm in}=0.5$,
    $\etaR=0.1$, comparing three- and four-source networks. The three-source
    networks outperform their four-source counterparts at fixed loss.}
    \label{fig:Fs34}
\end{figure}

Figures~\ref{fig:Fph34} and~\ref{fig:Fs34} compare the three- and four-source
networks directly. Before postselection ($\eta_{\rm in}=0.5$,
Fig.~\ref{fig:Fph34}), the three-source PBP source has the higher fidelity because
its heralding photons traverse fewer lossy stages. The ordering reverses for the
all-BSM benchmarks: the three-source chain contains the large vacuum contribution
in Eq.~\eqref{eq:psiBSM3}, whereas the four-source chain does not. After receiver
postselection ($\eta_{\rm in}=0.5$, $\etaR=0.1$,
Fig.~\ref{fig:Fs34}), the three-source networks again retain the higher
fidelity at fixed component loss. The value of the fourth source lies elsewhere:
it removes the three-source residual term in the ideal architecture and adds an
elementary link, trading additional loss for greater span (with the
transmissions of Sec.~\ref{sec:three}, $6L+2L'\approx44$~km compared with
33~km).

\subsection{Success probability and delivery rate}

\begin{figure}[tbp]
    \centering
    \includegraphics{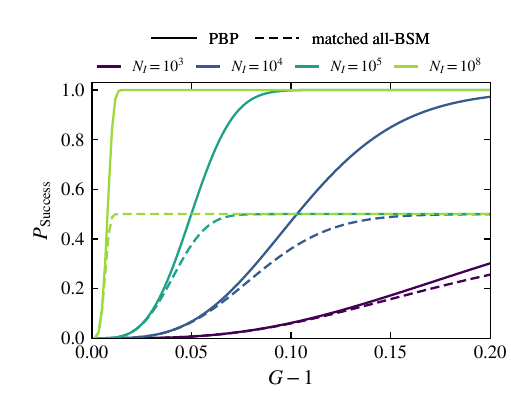}
    \caption{Success probability $P_{\rm Success}$ versus $G-1$ for
    $N_I=10^{3},10^{4},10^{5},10^{8}$ (bottom to top), four-source, lossless.
    Solid: \PBP; dashed: matched \allBSM\ chain. The \PBP\ network is
    asymptotically deterministic ($P_{\rm Success}\to1$); the \allBSM\ chain is
    bounded by $0.5$.}
    \label{fig:Psucc4}
\end{figure}

Figure~\ref{fig:Psucc4} compares the four-source success probability versus $G-1$
for several $N_I$ in the lossless case. Multiplexing drives the PBP probability
toward its unit Bell-fraction ceiling, so that the lossless four-source network is
asymptotically deterministic, while the all-BSM chain remains capped at $1/2$.
Loss increases the mode count needed to approach either ceiling; the PBP
advantage persists. For example, with effective heralding transmission
$\eta_{\rm in}=0.7$, $G-1=0.005$, and $N_I=5\times10^9$, the PBP success
probability is $0.75$, compared with $0.47$ for the all-BSM chain. Such mode
counts, like the $10^7$ of Sec.~\ref{sec:three}, lie far beyond a purely
spectral implementation and are used here to display the asymptotic behavior;
the four-source network requires several orders of magnitude more same-pulse
trials than the three-source network because its herald probability carries an
additional stage.

\begin{figure}[tbp]
    \centering
    \includegraphics{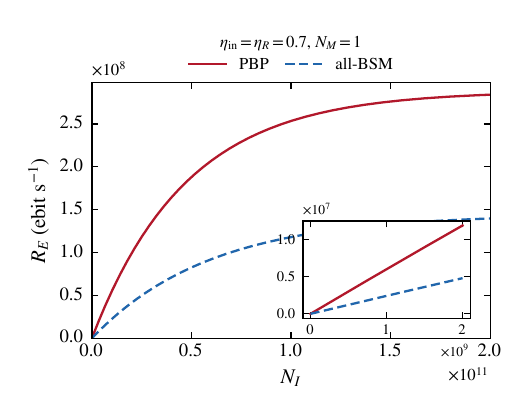}
    \caption{Entanglement-delivery rate $\RE$ (ebit/s) versus $N_I$, four-source,
    single-memory ($N_M=1$). Solid: \PBP; dashed: matched \allBSM\
    chain. Parameters: $\RP=10^{9}\,\mathrm{s}^{-1}$, $\eta_{\rm in}=0.7$,
    $\etaR=0.7$; $G-1=2.71\times10^{-3}$ (\PBP), $2.16\times10^{-3}$ (\allBSM).}
    \label{fig:RE4single}
\end{figure}

\begin{figure}[tbp]
    \centering
    \includegraphics{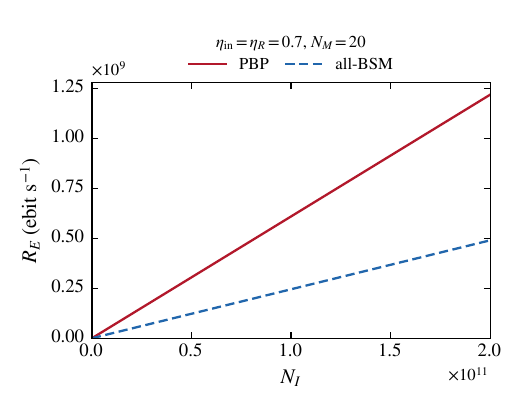}
    \caption{Entanglement-delivery rate $\RE$ (ebit/s) versus $N_I$, four-source,
    multi-memory ($N_M=20$). Solid: \PBP; dashed: matched \allBSM\
    chain. Parameters as in Fig.~\ref{fig:RE4single}.}
    \label{fig:RE4multi}
\end{figure}

Finally, Figs.~\ref{fig:RE4single} and~\ref{fig:RE4multi} give the four-source
delivery rate versus multiplexing for single-memory and $N_M=20$ operation, at
$\calF_S=0.99$, $\eta_{\rm in}=\etaR=0.7$, and
$\RP=10^{9}\,\mathrm{s}^{-1}$ [\PBP\ at $G-1=2.71\times10^{-3}$, \allBSM\ at
$2.16\times10^{-3}$]. With one receiver memory, the rates approach saturation
near $N_I\sim2\times10^{11}$ and the PBP rate is approximately $2$--$2.5$ times
the all-BSM rate. With $N_M=20$, both rates remain nearly linear over the plotted
range and their ratio stays close to $2.5$; at $N_I=2\times10^{11}$, PBP reaches
$1.22\times10^9$~ebit/s. Additional memories raise the aggregate rate and
postpone saturation. The mechanism is the same as in Sec.~\ref{sec:three}:
with one memory the all-BSM chain saturates at the rate set by its same-side
false heralds, which account for roughly half of its heralds, whereas with
$N_M=20$ both networks remain channel-limited over the plotted range.

The four-source construction therefore completes the complementary filtering
sequence: it removes the residual term of the minimal three-source transmitter and
produces an exact Bell state in the lossless limit. Its additional source and
measurement stage increase the loss and multiplexing burden, but the PBP advantage
over the matched all-BSM chain persists in both fidelity and delivery rate.

\section{Conclusion}
\label{sec:conclusion}

We have recast multipair contamination in cascaded entanglement swapping as a
class-covering problem. A conventional BSM rejects the same-rail,
same-polarization double-emission class $\mathcal C_4$ but cannot distinguish a
same-rail, mixed-polarization double emission from one
input from the desired one-photon-per-input sector. Entanglement swapping
provides more than one measurement location, and those measurements need not be
identical. Pairing the direct-basis BSM with an equatorial analyzer makes their
rejection zeros complementary, so a double emission from an internal source,
whose two rails feed the two analyzers, cannot pass both and cannot produce an
end-to-end herald. The same pairing also recovers a useful sector: the GM turns
$\mathcal C_3$ inter-source events rejected by the BSM into resolved
$\phi$-type heralds. For the balanced analyzers, this recovery exactly offsets
the GM's reduced $\mathcal C_1$ acceptance, preserving the ideal true-event
probability while suppressing false heralds.

Within the passive four-mode architecture considered here (PNR detection with a
two-bank acceptance rule), the balanced BSM and GM are the canonical
representatives of the resulting analyzer family. A common equatorial phase does not change the accepted
probabilities, loss dependence, or desired Bell component; a relative phase
between the rail analyzers produces only a known local Bell-state rotation. This
freedom can accommodate a chosen equatorial encoding or stable birefringent phase
without changing the filtering mechanism. The essential ingredient is therefore
a pair of measurements whose many-boson interference zeros cover complementary
error classes; which balanced multiport implements the second measurement is
secondary.

The minimal realization is the three-source BSM--GM cascade. It removes the
vacuum-dominated false-herald background of the matched BSM--BSM chain and
leaves only the higher-order term in Eqs.~\eqref{eq:psiS3} and
\eqref{eq:psiB3}. Adding a fourth source produces the symmetric BSM--GM--BSM
cascade, which removes that residual and heralds an exact Bell state in the
lossless limit. As shown in Sec.~\ref{sec:four} and
Appendix~\ref{appendix:fock_state_analysis}, every lossless alternating BSM--GM
chain with $N\geq4$ sources heralds a pure outer Bell state to all emission
orders. Photon-number conservation reduces all accepted events to only three
global pair-number histories, and the two unwanted histories are annihilated by
an internal source lying between complementary analyzers. Exhaustive enumeration
for $N=4$--$6$ independently verifies the result for every accepted record
sequence.

The all-orders Gaussian calculation shows that the advantage survives detector,
heralding-path, and propagation loss. In the lossless high-multiplexing limit,
the three-source success probability approaches unity at low brightness and the
four-source network is asymptotically deterministic, whereas the matched all-BSM
ceilings are
set by their false-herald fractions. Under the lossy operating points studied
here, PBP also provides the higher fidelity-matched delivery rate. Multiplexing
amplifies this advantage, which is already present in a single channel. Receiver
loss can raise a conditional fidelity when internal loss has admitted additional
higher-order histories; this is a reweighting of the loaded ensemble, it lowers
the loading probability, and it is absent for the ideal heralded state.

The improvement has a clear resource cost. All stages must herald in the same
channel and pump pulse, so a memory-free implementation requires many synchronous
trials; adding the fourth source further increases the loss and mode-count burden.
The three-source transmitter consequently gives the better fidelity at fixed component
loss, whereas the four-source transmitter trades that advantage for an additional
elementary link and exact lossless filtering. Combining frequency, time, and
spatial multiplexing can distribute the required mode count. More broadly, the
result suggests a design procedure for heralding networks: identify the unwanted
symmetry classes first, then choose analyzers whose interference zeros cover
them.

\begin{acknowledgments}
This work was partially supported by the National Science Foundation under Grant
No.~1941583.
\end{acknowledgments}

\appendix
\onecolumngrid
\makeatletter
\@addtoreset{table}{section}
\makeatother
\renewcommand{\thetable}{\Alph{section}\arabic{table}}

\section{Classification of complementary two-photon analyzers}
\label{app:analyzer_classification}

This Appendix classifies the two-operation solutions of
Eq.~\eqref{eq:intro_covering} within the measurement architecture used in the
main text. Each operation consists of a lossless passive four-mode unitary,
photon-number-resolving detection, and a two-bank acceptance rule requiring one
photon in each of two two-mode detector banks. The classification concerns local
acceptance at each station; it does not cover an arbitrary decoder that makes a
joint decision from the complete record pair of both stations.

\subsection{Two-bank analyzers}

Let the single-photon input space be
\begin{equation}
\mathcal H=\mathcal H_{\rm rail}\otimes\mathcal H_{\rm pol},
\qquad
(e_1,e_2,e_3,e_4)=(a_H,a_V,b_H,b_V).
\label{eq:classification_space}
\end{equation}
Pulling the two detector banks back through an arbitrary four-mode unitary
defines a Hermitian involution on this space,
\begin{equation}
Z=Z^\dagger,\qquad Z^2=I,\qquad \operatorname{tr}Z=0,
\label{eq:classification_involution}
\end{equation}
whose $+1$ and $-1$ eigenspaces are the two input-space bank subspaces. Unitary
rotations within either bank do not change the total acceptance probability and
only resolve that probability into different detector records. On the symmetric
two-photon space, the projector onto one photon in each bank is
\begin{equation}
\Pi_{\rm acc}(Z)=\frac{I-Z\otimes Z}{2}.
\label{eq:classification_projector}
\end{equation}
An entire class $\mathcal C_k$ is therefore rejected if and only if
\begin{equation}
(Z\otimes Z)\ket\psi=\ket\psi,
\qquad \ket\psi\in\mathcal C_k.
\label{eq:classification_dark}
\end{equation}

Consider first an operation that rejects $\mathcal C_4$. Applying
Eq.~\eqref{eq:classification_dark} to each square
$e_i^\dagger e_i^\dagger\ket0$ shows that every computational mode is an
eigenvector of $Z$:
\begin{equation}
Ze_i=s_i e_i,\qquad s_i\in\{+1,-1\}.
\label{eq:classification_signs}
\end{equation}
The trace condition requires two signs of each type. Up to an overall sign, the
three possible bank partitions are
\begin{align}
Z_{\rm pol}&=\operatorname{diag}(+,-,+,-),\nonumber\\
Z_{\rm rail}&=\operatorname{diag}(+,+,-,-),\nonumber\\
Z_{\rm diag}&=\operatorname{diag}(+,-,-,+).
\label{eq:classification_parities}
\end{align}
The diagonal partition $Z_{\rm diag}$ also rejects the desired class
$\mathcal C_1$ and is inadmissible under Eq.~\eqref{eq:intro_covering}; it is
$Z_{\rm pol}$ conjugated by a polarization flip in one rail, so admitting it
would add only this local relabeling of the BSM. The remaining two signatures are
\begin{equation}
\begin{array}{c|cccc}
 &\mathcal C_1&\mathcal C_2&\mathcal C_3&\mathcal C_4\\ \hline
Z_{\rm pol}&>0&>0&0&0\\
Z_{\rm rail}&>0&0&>0&0.
\end{array}
\label{eq:classification_parity_map}
\end{equation}
Thus any admissible operation that rejects $\mathcal C_4$ measures either
polarization parity or rail parity.

Next consider all operations that reject $\mathcal C_2$. Rejection of
$a_H^\dagger a_V^\dagger\ket0$ requires $Z$ to preserve the two-dimensional
rail-$a$ polarization subspace and to preserve the unordered pair of rays
$\{a_H,a_V\}$. Within that subspace a Hermitian involution can act only as
\begin{equation}
\pm I_2
\quad\text{or}\quad
X(\phi)=
\begin{pmatrix}
0&e^{-i\phi}\\ e^{i\phi}&0
\end{pmatrix}.
\label{eq:classification_block_options}
\end{equation}
The same result holds independently in rail $b$. Imposing
$\operatorname{tr}Z=0$ leaves, up to overall signs and bank relabeling,
\begin{align}
Z_{\rm eq}(\phi_a,\phi_b)
&=X(\phi_a)_a\oplus X(\phi_b)_b,
\label{eq:classification_zeq}\\
Z_{\rm rail}&=+I_a\oplus(-I_b).
\end{align}
The first operation measures an equatorial polarization basis independently in
each rail and has signature
\begin{equation}
\begin{array}{c|cccc}
 &\mathcal C_1&\mathcal C_2&\mathcal C_3&\mathcal C_4\\ \hline
Z_{\rm eq}&>0&0&>0&>0.
\end{array}
\label{eq:classification_equatorial_map}
\end{equation}

\subsection{Complete covering solutions}

At least one of the two operations must reject $\mathcal C_4$. If that operation
is $Z_{\rm pol}$, the second operation need only reject $\mathcal C_2$, and
Eqs.~\eqref{eq:classification_block_options}
and~\eqref{eq:classification_zeq} give
\begin{equation}
(Z_A,Z_B)=
\bigl(Z_{\rm pol},Z_{\rm eq}(\phi_a,\phi_b)\bigr)
\quad\text{or}\quad
(Z_{\rm pol},Z_{\rm rail}).
\label{eq:classification_branch_pol}
\end{equation}
If the first operation is instead $Z_{\rm rail}$, it already rejects both
$\mathcal C_2$ and $\mathcal C_4$, and the covering condition places no further
constraint on the second operation,
\begin{equation}
(Z_A,Z_B)=(Z_{\rm rail},Z_B),\qquad Z_B\ \text{arbitrary admissible}.
\label{eq:classification_branch_rail}
\end{equation}
Up to exchanging $A$ and $B$, overall signs, and detector-bank
labels, Eqs.~\eqref{eq:classification_branch_pol} and
\eqref{eq:classification_branch_rail} exhaust the solutions of the class-covering
condition.

The class constraints alone therefore admit solutions containing
$Z_{\rm rail}$. Such an operation has the original input rails as its two bank
subspaces: every accepted record identifies one mode from rail $a$ and one from
rail $b$. It preserves which-source information and implements a separable
measurement across the two inputs, so it cannot perform entanglement swapping on
two independent source pairs. Excluding these nonentangling branches leaves the
single physical structural family
\begin{equation}
\boxed{
Z_A=I_{\rm rail}\otimes \sigma_z,
\qquad
Z_B=X(\phi_a)_a\oplus X(\phi_b)_b .}
\label{eq:classification_physical_family}
\end{equation}
The two operations are therefore a direct-basis polarization-parity analyzer and
an equatorial polarization analyzer. This establishes the qualified uniqueness
claim of Secs.~\ref{sec:intro} and~\ref{sec:general}.

\subsection{Resolved records and balanced representatives}

Equation~\eqref{eq:classification_physical_family} fixes the detector-bank
subspaces but leaves arbitrary rotations within them. Up to output permutations
and phases, the full equatorial unitary family may be written
\begin{equation}
U_{\rm eq}=(W_+\oplus W_-)
\bigl[R(\phi_a)\oplus R(\phi_b)\bigr],
\qquad W_+,W_-\in U(2),
\label{eq:classification_unitary_family}
\end{equation}
where $R(\phi_r)$ maps $H/V$ in rail $r$ to the eigenvectors of
$X(\phi_r)$ and a fixed permutation groups the two equatorial detector banks.
The matrices $W_+$ and $W_-$ mix the rails independently within the two banks.

The resolved records determine when a class-level success is also an entangling
herald. For a polarization-parity record, let
$h=(h_a,h_b)$ and $v=(v_a,v_b)$ be the rail amplitudes of the detected $H$- and
$V$-bank modes. Its projection onto $\mathcal C_1$ is maximally entangled exactly
when
\begin{equation}
|h_a v_b|=|h_b v_a|\neq0.
\label{eq:classification_bsm_record}
\end{equation}
Requiring all four retained opposite-polarization records to satisfy this
condition gives balanced rail mixing in both banks, as in the conventional BSM.

For an equatorial record, let $A=(A_a,A_b)$ and $B=(B_a,B_b)$ be the rail
amplitudes of the detected modes in the two banks. Its two $\mathcal C_1$
amplitudes are proportional to
\begin{equation}
e^{i\phi_b}\Delta,
\qquad -e^{i\phi_a}\Delta,
\qquad
\Delta=A_aB_b-A_bB_a.
\label{eq:classification_gm_record}
\end{equation}
They have equal magnitude automatically, and the record is entangling whenever
$\Delta\neq0$. The phase difference $\phi_a-\phi_b$ only fixes the known local
phase of the heralded Bell state. The rail-symmetric choice
$\phi_a=\phi_b=\phi$ and $W_+=W_-=H_2$ gives the balanced family
$U_\phi$ in Eq.~\eqref{eq:Uphi_def}; its real member is the GM. Thus the
class constraints select the direct-parity/equatorial structure, while balance
and rail symmetry select the canonical BSM--GM implementation used here.

\section{Analysis of the three-source PBP source using Gaussian states}
\label{app:three}

We derive the Gaussian-state model of the three-source \PBP\ source under loss. The
source consists of three type-II Sagnac sources; for source $i$ the output modes are
$\hat a_{iH},\hat a_{iV},\hat b_{iH},\hat b_{iV}$. Modes $b_1$ (source~1) and $a_2$
(source~2) interfere at the BSM, and modes $b_2$ (source~2) and $a_3$ (source~3)
interfere in the GM; the projected outer modes are $a_1$ and $b_3$.
We first combine sources~1 and~2 with the BSM, then combine the result with source~3
via the GM.

Each Sagnac source is built from two independent TMSVs generating pairs in the
cross-polarized mode pairs $(a_{iH},b_{iV})$ and $(a_{iV},b_{iH})$; source $i$ is
associated with TMSVs $2i-1$ and $2i$. Its output state is
\begin{equation}
    \ket{\psi}_i = \sum_{n_i=0}^{\infty}\sum_{m_i=0}^{\infty}(-1)^{m_i}
    \sqrt{P_{n_i}P_{m_i}}\;\ket{n_i}_{a_{iH}}\ket{m_i}_{a_{iV}}\ket{m_i}_{b_{iH}}\ket{n_i}_{b_{iV}},
    \label{eq:A1}
\end{equation}
with $P_n=(G-1)^n/G^{n+1}$. Combining sources~1 and~2, the outer modes are
$O=(a_{1H},a_{1V},b_{2H},b_{2V})$ and the BSM-interfering inner modes are
$I=(b_{1H},b_{1V},a_{2H},a_{2V})$, with complex variables $z_O$ and $z_I$.

\subsection{Before postselection}

The anti-normally-ordered characteristic function (ANOCF) of the density operator
$\hat\rho_L$ of the combined 1--2 source is
\begin{equation}
    \chi_A^{\rho_L}(\bm{z}) = \Tr\!\left[\hat\rho_{L}\,e^{-\bm{z}^{\dagger}\hat{\bm a}}\,e^{\hat{\bm a}^{\dagger}\bm{z}}\right],
    \label{eq:A_anocf_def}
\end{equation}
with $\bm z^\dagger=[\bm z_O^\dagger\;\bm z_I^\dagger]$ and
$\hat{\bm a}^\dagger=[\hat{\bm a}_O^\dagger\;\hat{\bm a}_I^\dagger]$, and
\begin{equation}
    \hat\rho_{L} = \int \frac{d^{16}\bm{z}}{\pi^{8}}\,\chi_A^{\rho_{L}}(\bm{z})\,e^{-\hat{\bm a}^{\dagger}\bm{z}}\,e^{\bm{z}^{\dagger}\hat{\bm a}}.
    \label{eq:A_inverse}
\end{equation}
Using \Eq{eq:A1},
\begin{equation}
    \chi_A^{\rho_{L}}(\bm{z}) = \exp\!\Big[-G\,\bm{z}^{\dagger}\bm{z}+2\sqrt{G(G-1)}\,\Re\!\big(\bm{z}_{O}^{T}K_{L}\bm{z}_{I}\big)\Big],
    \label{eq:A_chi_raw}
\end{equation}
where for type-II interactions
\begin{equation}
    K_{L} = \begin{pmatrix}0&1&0&0\\-1&0&0&0\\0&0&0&-1\\0&0&1&0\end{pmatrix}.
    \label{eq:A_KL}
\end{equation}
To obtain the conditioned output states we (i) apply the 50:50 beam-splitter
transformation of the BSM, (ii) apply detector loss, (iii) invert to the density
operator, (iv) condition on the detected record, and (v) recompute the
conditioned characteristic function; we then repeat with source~3 via the GM.

Separating real and imaginary parts, $\bm z=\bm z^{(R)}+i\bm z^{(I)}$, and using
that $K_L$ is real,
\begin{equation}
    \chi_A^{\rho_L}(\bm z)
    =\exp\!\Big[-\bm z^{(R)T}\tilde\Lambda^{(R)}\bm z^{(R)}/2\Big]
     \exp\!\Big[-\bm z^{(I)T}\tilde\Lambda^{(I)}\bm z^{(I)}/2\Big],
    \label{eq:A_split}
\end{equation}
with
\begin{equation}
    \tilde{\Lambda}^{(R)} = \begin{bmatrix}2G\,I_4 & -2\sqrt{G(G-1)}\,K_{L}\\ -2\sqrt{G(G-1)}\,K_{L}^{T} & 2G\,I_4\end{bmatrix},\quad
    \tilde{\Lambda}^{(I)} = \begin{bmatrix}2G\,I_4 & 2\sqrt{G(G-1)}\,K_{L}\\ 2\sqrt{G(G-1)}\,K_{L}^{T} & 2G\,I_4\end{bmatrix}.
    \label{eq:A_Lambdas}
\end{equation}
The balanced BSM maps $I\to I'$ by
\begin{equation}
    \begin{pmatrix}\hat b_{1H'}\\\hat b_{1V'}\\\hat a_{2H'}\\\hat a_{2V'}\end{pmatrix}
    =B\begin{pmatrix}\hat b_{1H}\\\hat b_{1V}\\\hat a_{2H}\\\hat a_{2V}\end{pmatrix},
    \quad B=B^{T}=\frac{1}{\sqrt2}\begin{pmatrix}-1&0&1&0\\0&-1&0&1\\1&0&1&0\\0&1&0&1\end{pmatrix},
    \label{eq:A_BS}
\end{equation}
giving the cross-correlation matrix
\begin{equation}
    K'_{L}=K_{L}B^{T}=\frac{1}{\sqrt2}\begin{pmatrix}0&-1&0&1\\1&0&-1&0\\0&-1&0&-1\\1&0&1&0\end{pmatrix},\quad K'_L(K'_L)^T=I_4.
    \label{eq:A_KLp}
\end{equation}
The BSM outputs enter four PNR detectors of efficiency $\etaT$, modeled as
$\hat a''_\mu=\sqrt{\etaT}\,\hat a'_\mu+\sqrt{1-\etaT}\,\hat v_\mu$; a pure-loss
channel transforms the ANOCF as
$\chi_A^{\rho''_L}(\bm z_O,\bm z_{I''})=e^{-(1-\etaT)\bm z_{I''}^\dagger\bm z_{I''}}\chi_A^{\rho'_L}(\bm z_O,\sqrt{\etaT}\,\bm z_{I''})$,
yielding
\begin{equation}
    \tilde\Lambda''^{(R)}=\begin{bmatrix}2G\,I_4 & -2\sqrt{\etaT G(G-1)}\,K'_L\\ -2\sqrt{\etaT G(G-1)}\,(K'_L)^T & 2[\etaT(G-1)+1]\,I_4\end{bmatrix},
    \label{eq:A_Lpp}
\end{equation}
and its imaginary counterpart with the sign of the off-diagonal block reversed. To
allow analytic polynomial-weighted inverse transforms we use the inverses
\begin{equation}
    \Lambda''^{(R)}=(\tilde\Lambda''^{(R)})^{-1}=\begin{bmatrix}\frac{[\etaT(G-1)+1]I_4}{2G} & \frac{\sqrt{\etaT G(G-1)}}{2G}K'_L\\ \frac{\sqrt{\etaT G(G-1)}}{2G}(K'_L)^T & \frac12 I_4\end{bmatrix},
    \label{eq:A_Lppinv}
\end{equation}
(and its imaginary counterpart), so that the ANOCF factorizes into Gaussian marginal
and conditional densities. With $\det\Lambda''^{(R)}=\det\Lambda''^{(I)}=1/256G^4$,
\begin{equation}
    \chi_A^{\rho''_L}(\bm z)=\Big(\frac{\pi^8}{G^4}\Big)\,
    p_{Z_O^{(R)}}\,p_{Z_{I''}^{(R)}|Z_O^{(R)}}\,
    p_{Z_O^{(I)}}\,p_{Z_{I''}^{(I)}|Z_O^{(I)}},
    \label{eq:A_factorized}
\end{equation}
where the outer marginals are zero-mean Gaussians with covariance
$[\etaT(G-1)+1]I_4/2G$. Conditioning on the BSM record
$\delta_L\in\{0011,1100,1001,0110\}$ (ordering
$(b''_{1H},b''_{1V},a''_{2H},a''_{2V})$), the unnormalized density operator for
$(1001)$ is
\begin{equation}
    \tilde\rho_{O|(1001)}=\frac{[\etaT(G-1)]^2}{[\etaT(G-1)+1]^6}\int\frac{d^8 z_O}{\pi^4}
    e^{-z_O^\dagger z_O/N_S}
    \Big[1-\tfrac{|z_{a1H}-z_{b2H}|^2}{2N_S}\Big]
    \Big[1-\tfrac{|z_{a1V}+z_{b2V}|^2}{2N_S}\Big]
    e^{-\hat a_O^\dagger z_O}e^{z_O^\dagger\hat a_O},
    \label{eq:A_rho1001}
\end{equation}
with $N_S=[\etaT(G-1)+1]/G$ and prefactor equal to the $(1001)$ generation
probability (equal for all patterns). Its ANOCF is
\begin{equation}
    \chi_A^{\rho_{O|(1001)}}(\bm z_O)=\frac{[\etaT(G-1)]^2}{[\etaT(G-1)+1]^6}
    e^{-z_O^\dagger z_O/N_S}
    \Big[1-\tfrac{|z_{a1H}-z_{b2H}|^2}{2N_S}\Big]
    \Big[1-\tfrac{|z_{a1V}+z_{b2V}|^2}{2N_S}\Big].
    \label{eq:A_chi1001}
\end{equation}

We now combine with source~3 via the GM.
Source~3 has ANOCF
\begin{equation}
    \chi_A^{\rho_3}(z_{O_3},z_{I_3})=\exp\!\Big[-G(z_{O_3}^\dagger z_{O_3}+z_{I_3}^\dagger z_{I_3})+2\sqrt{G(G-1)}\,\Re(z_{O_3}^T K_2 z_{I_3})\Big],
    \quad K_2=\begin{pmatrix}0&1\\-1&0\end{pmatrix},
    \label{eq:A_chi3}
\end{equation}
with $O_3=(b_{3H},b_{3V})$, $I_3=(a_{3H},a_{3V})$. The joint pre-GM state is
$\tilde\rho_{\rm in}=\tilde\rho_{O|(1001)}\otimes\rho_3$. The GM acts
on the inner modes $(b_2,a_3)$ by the transform
$U_\phi=B_{\rm rail}\otimes\Rphi$ of Eq.~\eqref{eq:Uphi_def}. In the input ordering
$(b_{2H},b_{2V},a_{3H},a_{3V})$,
\begin{equation}
    H(\phi)=U_0\,\mathrm{diag}(1,e^{i\phi},1,e^{i\phi})=\frac12\begin{pmatrix}
    1&e^{i\phi}&1&e^{i\phi}\\1&-e^{i\phi}&1&-e^{i\phi}\\1&e^{i\phi}&-1&-e^{i\phi}\\1&-e^{i\phi}&-1&e^{i\phi}\end{pmatrix},
    \qquad
    H\equiv H(0)=\frac12\begin{pmatrix}1&1&1&1\\1&-1&1&-1\\1&1&-1&-1\\1&-1&-1&1\end{pmatrix},
    \label{eq:A_H}
\end{equation}
i.e., the real-Hadamard Green Machine $H$ preceded by a phase $e^{i\phi}$ on the
vertical inputs $(b_{2V},a_{3V})$. Every entry of $H(\phi)$ has modulus $1/2$;
varying $\phi$ changes only the input phases. We carry the general $H(\phi)$ and
specialize to $\phi=0$ for the numerics. Writing the GM output variables
$u$ and defining
$\alpha_H=\tfrac{\sqrt{\etaG}}{2}(u_1+u_2+u_3+u_4)$,
$\alpha_V=e^{i\phi}\tfrac{\sqrt{\etaG}}{2}(u_1-u_2+u_3-u_4)$,
$\beta_H=\tfrac{\sqrt{\etaG}}{2}(u_1+u_2-u_3-u_4)$,
$\beta_V=e^{i\phi}\tfrac{\sqrt{\etaG}}{2}(u_1-u_2-u_3+u_4)$, the lossy GM-stage
ANOCF is
\begin{equation}
\begin{split}
\tilde\chi_{\rm GM,A}(\bm z_G,\bm u)=C_L\,
&\exp\!\Big[-\tfrac{|z_{a1H}|^2+|z_{a1V}|^2+|\alpha_H|^2+|\alpha_V|^2}{N_S}\Big]
\Big(1-\tfrac{|z_{a1H}-\alpha_H|^2}{2N_S}\Big)\Big(1-\tfrac{|z_{a1V}+\alpha_V|^2}{2N_S}\Big)\\
&\times\exp\!\Big[-G(|\beta_H|^2+|\beta_V|^2+|z_{b3H}|^2+|z_{b3V}|^2)\Big]\\
&\times\exp\!\Big[2\sqrt{G(G-1)}\,\Re(\beta_V z_{b3H}-\beta_H z_{b3V})\Big]
e^{-(1-\etaG)u^\dagger u},
\end{split}
\label{eq:A_chiGM}
\end{equation}
with $C_L=[\etaT(G-1)]^2/[\etaT(G-1)+1]^6$ and
$\bm z_G^\dagger=[z_{a1H}^*\,z_{a1V}^*\,z_{b3H}^*\,z_{b3V}^*]$. Conditioning on the
GM record $(1001)$ inserts the Laguerre factors $(1-|u_1|^2)(1-|u_4|^2)$;
integrating over $u$ gives the conditioned outer ANOCF
$\tilde\chi_{\rm out,A}(\bm z_O)$, from which the outer density operator \(\tilde\rho_{\rm out|(1001)}\) follows by
inverse transform.

The phase $\phi$ enters \Eq{eq:A_chiGM} only through $\alpha_V\to e^{i\phi}\alpha_V$
and $\beta_V\to e^{i\phi}\beta_V$; $\alpha_H,\beta_H$ are unchanged. Every kernel
below contains equal total powers of this common phase and its conjugate after
the Gaussian integration. Consequently $P_{\rm gen}$, $P_{\rm Bell}$
[\Eq{eq:A_PBell}], and all reported fidelities and rates are $\phi$-independent.
In the lossless Fock state the desired Bell doublet is itself unchanged; $\phi$
only changes its phase relative to orthogonal multiphoton sectors. We set
$\phi=0$ hereafter when evaluating the figures of merit.

The Bell overlaps
$P_{\psi^\pm}\equiv\bra{\psi^\pm}\tilde\rho_{\rm out|(1001)}\ket{\psi^\pm}$ and
$P_{\phi^\pm}$ are computed from
\begin{equation}
P_{\psi^{\pm}}=\int d^8 z_G\,\mathcal{M}(z_G)\,K_{\psi^{\pm}}(z_G),\qquad
P_{\phi^{\pm}}=\int d^8 z_G\,\mathcal{M}(z_G)\,K_{\phi^{\pm}}(z_G),
\label{eq:A_bell}
\end{equation}
with kernels
\begin{align}
K_{\psi^{\pm}}(z_G)&=\tfrac12\Big[(1-|z_{a1H}|^2)(1-|z_{b3V}|^2)\pm2\Re(z_{a1H}z_{a1V}^*z_{b3V}z_{b3H}^*)+(1-|z_{a1V}|^2)(1-|z_{b3H}|^2)\Big],\\
K_{\phi^{\pm}}(z_G)&=\tfrac12\Big[(1-|z_{a1H}|^2)(1-|z_{b3H}|^2)\pm2\Re(z_{a1H}z_{a1V}^*z_{b3H}z_{b3V}^*)+(1-|z_{a1V}|^2)(1-|z_{b3V}|^2)\Big],
\label{eq:A_kernels}
\end{align}
and with $\mathcal{M}(z_G)$ and the auxiliary integral $W(\bm z_G)$ defined by
\begin{equation}
\begin{split}
W(\bm z_G)=\int\frac{d^8\bm u}{\pi^4}\,
&\exp\!\Big[-\tfrac{|\alpha_H|^2+|\alpha_V|^2}{N_S}\Big]\exp\!\Big[-\tfrac{|\beta_H|^2+|\beta_V|^2}{N_C}\Big]
\Big(1-\tfrac{|z_{a1H}-\alpha_H|^2}{2N_S}\Big)\Big(1-\tfrac{|z_{a1V}+\alpha_V|^2}{2N_S}\Big)\\
&\times(1-|u_1|^2)(1-|u_4|^2)\,\exp\!\Big[2\sqrt{G(G-1)}\,\Re(\beta_V z_{b3H}-\beta_H z_{b3V})\Big]e^{-(1-\etaG)u^\dagger u},
\end{split}
\label{eq:A_W}
\end{equation}
\begin{equation}
\mathcal{M}(z_G)\equiv\frac{1}{\pi^4}\frac{[\etaT(G-1)]^2}{[\etaT(G-1)+1]^{6}}
\exp\!\Big[-\tfrac{|z_{a1H}|^2+|z_{a1V}|^2}{N_S}\Big]\exp\!\Big[-\tfrac{|z_{b3H}|^2+|z_{b3V}|^2}{N_C}\Big]W(\bm z_G).
\label{eq:A_M}
\end{equation}
The generation probability is
$P_{\rm gen}=\Tr(\tilde\rho_{\rm out|(1001)})=\tilde\chi_{\rm out,A}(0)$, and
$P_{\rm Bell}=P_{\psi^-}+P_{\psi^+}+P_{\phi^-}+P_{\phi^+}$, i.e.,
\begin{equation}
P_{\rm Bell}=\int d^8 z_G\,\mathcal{M}(z_G)\big(2-|z_{a1H}|^2-|z_{a1V}|^2\big)\big(2-|z_{b3H}|^2-|z_{b3V}|^2\big).
\label{eq:A_PBell}
\end{equation}
The Bell fraction and target-state fidelity before postselection are
\begin{equation}
B_{\rm Ph}=\frac{P_{\rm Bell}}{P_{\rm gen}},\qquad
\calF_{\rm Ph}=\frac{P_{\psi^-}}{P_{\rm gen}}.
\label{eq:A_frac_fid}
\end{equation}

\subsection{After postselection}

We include QTX-to-QRX propagation loss, folding all channel, coupling, and
conversion losses into $\etaR$. With $A\equiv a_1$ (Alice), $B\equiv b_3$ (Bob), and
$\bm z_{AB}^\dagger=[z_{AH}^*\,z_{AV}^*\,z_{BH}^*\,z_{BV}^*]$,
\begin{equation}
\tilde\chi_{AB,A}^{(\etaR)}(z_{AB})=\tilde\chi_{\rm out,A}(\sqrt{\etaR}\,z_{AB})\,e^{-(1-\etaR)z_{AB}^\dagger z_{AB}},
\label{eq:A_etaR}
\end{equation}
from which the receiver-side singlet probability $P_{\psi^-,{\rm load}}$ and the
other Bell overlaps follow as in \Eq{eq:A_bell}. For ideal memories the loaded density
operator is
\begin{equation}
\tilde\rho_L^{(\etaR)}=\tilde\rho_{AB}^{(\etaR)}
-(\Pi_{0A}\!\otimes\!I_B)\tilde\rho_{AB}^{(\etaR)}(\Pi_{0A}\!\otimes\!I_B)
-(I_A\!\otimes\!\Pi_{0B})\tilde\rho_{AB}^{(\etaR)}(I_A\!\otimes\!\Pi_{0B})
+(\Pi_{0A}\!\otimes\!\Pi_{0B})\tilde\rho_{AB}^{(\etaR)}(\Pi_{0A}\!\otimes\!\Pi_{0B}),
\label{eq:A_loaded}
\end{equation}
with $\Pi_{0A}=\ket{0}_A\!\bra{0}$, $\Pi_{0B}=\ket{0}_B\!\bra{0}$, and the loadable
probability
\begin{equation}
P_{\rm Loadable}=\tilde\chi_{AB,A}^{(\etaR)}(0)
-\int\frac{d^4 z_A}{\pi^2}\tilde\chi_{AB,A}^{(\etaR)}(z_A,0)
-\int\frac{d^4 z_B}{\pi^2}\tilde\chi_{AB,A}^{(\etaR)}(0,z_B)
+\int\frac{d^8 z_{AB}}{\pi^4}\tilde\chi_{AB,A}^{(\etaR)}(z_{AB}).
\label{eq:A_Ploadable}
\end{equation}
The spin--spin fidelity and delivery rate follow from
\Eqs{eq:Fs}{eq:RE_general}. For the matched \allBSM\ chain, the same
steps apply with the GM transform $H(\phi)$ replaced by a beam-splitter transform.

\section{Analysis of the four-source PBP source using Gaussian states}
\label{app:four}

We combine four Sagnac sources with two BSMs and one GM. Sources~1--2
give, for BSM record $\delta_L$,
\begin{equation}
    \chi_A^{\rho_{X_L|\delta_L}}(\bm z_{X_L})=\frac{[\etaT(G-1)]^2}{[\etaT(G-1)+1]^6}
    e^{-z_{X_L}^\dagger z_{X_L}/N_S}
    \Big[1-\tfrac{|z_{a1H}+\sigma_H^{(L)}z_{b2H}|^2}{2N_S}\Big]
    \Big[1-\tfrac{|z_{a1V}+\sigma_V^{(L)}z_{b2V}|^2}{2N_S}\Big],
    \label{eq:B_left}
\end{equation}
with $X_L=(a_{1H},a_{1V},b_{2H},b_{2V})$ and signs
\begin{equation}
\begin{array}{c|cc}
\delta_L & \sigma_H^{(L)} & \sigma_V^{(L)}\\\hline
0011 & -1 & -1\\ 1100 & +1 & +1\\ 1001 & -1 & +1\\ 0110 & +1 & -1
\end{array}.
\label{eq:B_signs}
\end{equation}
Sources~3--4 give the analogous $\chi_A^{\rho_{X_R|\delta_R}}$ with
$X_R=(a_{3H},a_{3V},b_{4H},b_{4V})$ and the same sign table for $\delta_R$. The
joint state factorizes,
$\tilde\rho_{\rm PBP}=\tilde\rho_{X_L|\delta_L}\otimes\tilde\rho_{X_R|\delta_R}$,
with total pattern probability
$P_{\delta_L\delta_R}=[\etaT(G-1)]^4/[\etaT(G-1)+1]^{12}$. Partitioning into outer
modes $(a_1,b_4)$ and the central modes $(b_2,a_3)$ entering the GM, and
applying the GM transform
$U_\phi=U_0\,\mathrm{diag}(1,e^{i\phi},1,e^{i\phi})$ on the vertical inputs
$(b_{2V},a_{3V})$ (so $\bm z_C=U_\phi\bm z_d$; $\phi=0$ is the real Green Machine),
the ANOCF becomes
\begin{equation}
    \chi_A^{\rm PBP}(\bm z_O,\bm z_d)=\frac{[\etaT(G-1)]^4}{[\etaT(G-1)+1]^{12}}\,
    e^{-(z_O^\dagger z_O+z_d^\dagger z_d)/N_S}\prod_{i=1}^4\Big[1-\tfrac{|f_i|^2}{2N_S}\Big],
    \label{eq:B_chi}
\end{equation}
with (at $\phi=0$)
\begin{align}
f_1&=z_{a1H}-\tfrac12(z_{d1}+z_{d2}+z_{d3}+z_{d4}), &
f_2&=z_{a1V}+\tfrac12(z_{d1}-z_{d2}+z_{d3}-z_{d4}),\\
f_3&=\tfrac12(z_{d1}+z_{d2}-z_{d3}-z_{d4})-z_{b4H}, &
f_4&=\tfrac12(z_{d1}-z_{d2}-z_{d3}+z_{d4})+z_{b4V}.
\label{eq:B_f}
\end{align}
For general $\phi$ the vertical combinations $f_2,f_4$ carry a factor $e^{i\phi}$ on
their $z_d$ part; as in Appendix~\ref{app:three} this leaves every modulus-based
figure of merit unchanged. For the rail-symmetric family this common phase does
not rotate the desired Bell component. Applying GM output loss $\etaG$
($f_k''=\sqrt{\etaG}f_k+\sqrt{1-\etaG}v_k$) gives
\begin{equation}
    \chi_A^{\rho_O}=\frac{[\etaT(G-1)]^4}{[\etaT(G-1)+1]^{12}}
    \exp\!\Big[-\tfrac{z_O^\dagger z_O}{N_S}-\tfrac{z_{d''}^\dagger z_{d''}}{N_S'}\Big]
    \prod_{i=1}^4\Big[1-\tfrac{|f_i(z_O,\sqrt{\etaG}z_{d''})|^2}{2N_S}\Big],
    \label{eq:B_chi_loss}
\end{equation}
with $N_S'=(\etaG/N_S+(1-\etaG))^{-1}$. Projecting onto a GM record
via Laguerre factors $L_{n_k}(|z_{d_k''}|^2)$ [$L_0=1$, $L_1(x)=1-x$; e.g.
$(1-|z_{d1''}|^2)(1-|z_{d4''}|^2)$ for $(1001)$] yields the conditioned outer
density operator, and the Bell overlaps follow as
\begin{equation}
\begin{split}
O'_{\psi^{\pm}}\equiv\bra{\psi^\pm}\tilde\rho_{O|(1001)}\ket{\psi^\pm}
=\frac{[\etaT(G-1)]^4}{[\etaT(G-1)+1]^{12}}\int\frac{d^8\zeta_O}{\pi^4}
&\,e^{-\zeta_O^\dagger\zeta_O/N_S}\,W(\zeta_O)\\
\times\tfrac12\Big[(1-|\zeta_{a1H}|^2)(1-|\zeta_{b4V}|^2)
&\pm2\Re(\zeta_{a1H}\zeta_{a1V}^*\zeta_{b4V}\zeta_{b4H}^*)+(1-|\zeta_{a1V}|^2)(1-|\zeta_{b4H}|^2)\Big],
\end{split}
\label{eq:B_overlap}
\end{equation}
and analogously for $O'_{\phi^\pm}$, with
\begin{equation}
W(\zeta_O)=\int\frac{d^8\zeta_{d''}}{\pi^4}\,e^{-\zeta_{d''}^\dagger\zeta_{d''}/N_S'}
(1-|\zeta_{d1''}|^2)(1-|\zeta_{d4''}|^2)\prod_{i=1}^4\Big[1-\tfrac{|f_i(\zeta_O,\sqrt{\etaG}\zeta_{d''})|^2}{2N_S}\Big].
\label{eq:B_Wfour}
\end{equation}
The Bell fraction and fidelity before postselection are obtained as in
Appendix~\ref{app:three}. Including QTX-to-QRX loss $\etaR$ on the four outputs,
$\chi_A^{\tilde\rho_{AB|(1001)}^{(\etaR)}}(\zeta_{AB})=e^{-(1-\etaR)\zeta_{AB}^\dagger\zeta_{AB}}\chi_A^{\tilde\rho_{O|(1001)}}(\sqrt{\etaR}\,\zeta_{AB})$,
one obtains the receiver-side Bell probabilities [with
$N_R=(\etaR/N_S+(1-\etaR))^{-1}$], the loaded density operator [as in
\Eq{eq:A_loaded}], and the loadable probability [as in \Eq{eq:A_Ploadable}]. The
postselected figures of merit follow as in Appendix~\ref{app:three}. For the
matched \allBSM\ chain, the GM transform is replaced throughout by a
beam-splitter transform.

\section{Fock-state analysis of the three- and four-source architectures}
\label{appendix:fock_state_analysis}

This Appendix derives the output state in the ideal, lossless case for one spectral channel. The three-source configuration is presented first. The four-source configuration is then summarized by reusing the same source, BSM, and Green Machine transformations.

\subsection{Three-source configuration}

Each type-II Sagnac source consists of two independent TMSVs that generate photon pairs in the cross-polarized mode pairs \((a_{iH},b_{iV})\) and \((a_{iV},b_{iH})\). For sources 1 and 2,
\(
n_1:(a_{1H},b_{1V}),\quad
n_2:(a_{1V},b_{1H}),\quad
n_3:(a_{2H},b_{2V}),\quad
n_4:(a_{2V},b_{2H}).
\)
Their dual-rail output states are

\begin{equation}
    \ket{\psi}_1
    =
    \sum_{n_1=0}^{\infty}\sum_{n_2=0}^{\infty}
    (-1)^{n_2}
    \sqrt{P_{n_1}P_{n_2}}\;
    \ket{n_1}_{a_{1H}}
    \ket{n_2}_{a_{1V}}
    \ket{n_2}_{b_{1H}}
    \ket{n_1}_{b_{1V}},
    \label{equation_A3}
\end{equation}

\begin{equation}
    \ket{\psi}_2
    =
    \sum_{n_3=0}^{\infty}\sum_{n_4=0}^{\infty}
    (-1)^{n_4}
    \sqrt{P_{n_3}P_{n_4}}\;
    \ket{n_3}_{a_{2H}}
    \ket{n_4}_{a_{2V}}
    \ket{n_4}_{b_{2H}}
    \ket{n_3}_{b_{2V}}.
    \label{equation_A4}
\end{equation}

The 50:50 balanced beam-splitter transformation associated with the first BSM is

\begin{equation}
    \begin{pmatrix}
        \hat{b}_{1H'}^\dagger \\
        \hat{b}_{1V'}^\dagger \\\hat{a}_{2H'}^\dagger\\
        
    \hat{a}_{2V'}^\dagger
    \end{pmatrix}
    =
    \frac{1}{\sqrt{2}}
    \begin{pmatrix}
        1  & 0 & 1& 0 \\
        0 & 1&0&1\\
        1 & 0&-1&0\\
        0 & 1&0&-1
    \end{pmatrix}
     \begin{pmatrix}
        \hat{b}_{1H}^\dagger \\
        \hat{b}_{1V}^\dagger \\\hat{a}_{2H}^\dagger\\
        
    \hat{a}_{2V}^\dagger
    \end{pmatrix}.
\end{equation}
This sign convention differs from that of Eq.~\eqref{eq:A_BS} by local phases,
which do not affect the heralded states. After number-resolving detection, the
unnormalized conditional state on the remaining modes is

\begin{equation}
    \begin{aligned}
        \ket{\psi'}_{12}
         & =
        \hat{\Pi}^{(i_1 j_1 k_1 l_1)} \, \hat{U}_{B_1} \ket{\psi}_{12}
        \\[4pt]
         & =
        \sqrt{i_1!\,j_1!\,k_1!\,l_1!}
        \sum_{\substack{n_1,n_2,n_3,n_4=0 \\ q_1,q_2,q_3,q_4}}^\infty
        \frac{(-1)^{\,n_2+n_4+q_1+q_3}}{n_1!\,n_2!\,n_3!\,n_4!}
        \frac{\sqrt{P_{n_1}P_{n_2}P_{n_3}P_{n_4}}}{\sqrt{2^{\,n_1+n_2+n_3+n_4}}}
        \\[4pt]
         & \quad\times
        \binom{n_2}{q_1}\binom{n_3}{q_2}\binom{n_1}{q_3}\binom{n_4}{q_4}
        \\[4pt]
         & \quad\times
        \delta_{i_1,\;q_1+q_2}\,
        \delta_{j_1,\;q_3+q_4}\,
        \delta_{k_1,\;n_2+n_3-q_1-q_2}\,
        \delta_{l_1,\;n_1+n_4-q_3-q_4}
        \\[4pt]
         & \quad\times
        (\hat a_{1H}^\dagger)^{n_1}
        (\hat a_{1V}^\dagger)^{n_2}
        (\hat b_{2H}^\dagger)^{n_4}
        (\hat b_{2V}^\dagger)^{n_3}
        \ket{0}.
    \end{aligned}
    \label{equation_A9}
\end{equation}

For source 3,
\(
n_5:(a_{3H},b_{3V}),\qquad
n_6:(a_{3V},b_{3H}).
\)
In creation-operator form,

\begin{equation}
    \begin{aligned}
        \ket{\psi}_3
        =
        \sum_{n_5,n_6=0}^{\infty}
        \frac{(-1)^{n_6}\sqrt{P_{n_5}P_{n_6}}}{n_5!\,n_6!}\,
        (\hat a_{3H}^\dagger)^{n_5}
        (\hat a_{3V}^\dagger)^{n_6}
        (\hat b_{3H}^\dagger)^{n_6}
        (\hat b_{3V}^\dagger)^{n_5}
        \ket{0}.
    \end{aligned}
    \label{equation_A11}
\end{equation}

The Green Machine acting on modes \(b_2\) and \(a_3\) is

\begin{equation}
    \begin{pmatrix}
        \hat{b}_{2H'}^\dagger \\
        \hat{b}_{2V'}^\dagger \\
        \hat{a}_{3H'}^\dagger \\
        \hat{a}_{3V'}^\dagger
    \end{pmatrix}
    =
    \frac{1}{2}
    \begin{pmatrix}
        1 & 1 & 1 & 1 \\
        1 & -1 & 1 & -1 \\
        1 & 1 & -1 & -1 \\
        1 & -1 & -1 & 1
    \end{pmatrix}
    \begin{pmatrix}
        \hat{b}_{2H}^\dagger \\
        \hat{b}_{2V}^\dagger \\
        \hat{a}_{3H}^\dagger \\
        \hat{a}_{3V}^\dagger
    \end{pmatrix}.
    \label{equation_A12_new}
\end{equation}

Applying the Green Machine projection gives the following unnormalized state on the outer modes \(a_1\) and \(b_3\):

\begin{equation}
    \begin{aligned}
        \ket{\psi}
         & =
        \sqrt{i_1!\,j_1!\,k_1!\,l_1!\;i_2!\,j_2!\,k_2!\,l_2!}
        \\[4pt]
         & \quad\times
        \sum_{\substack{n_1,n_2,n_3,n_4=0 \\ q_1,q_2,q_3,q_4}}^\infty
        \frac{(-1)^{\,n_2+n_4+q_1+q_3}}{n_1!\,n_2!\,n_3!\,n_4!}
        \frac{\sqrt{P_{n_1}P_{n_2}P_{n_3}P_{n_4}}}{\sqrt{2^{\,n_1+n_2+n_3+n_4}}}
        \binom{n_2}{q_1}\binom{n_3}{q_2}\binom{n_1}{q_3}\binom{n_4}{q_4}
        \\[4pt]
         & \quad\times
        \delta_{i_1,\;q_1+q_2}\,
        \delta_{j_1,\;q_3+q_4}\,
        \delta_{k_1,\;n_2+n_3-q_1-q_2}\,
        \delta_{l_1,\;n_1+n_4-q_3-q_4}
        \\[4pt]
         & \quad\times
        \frac{1}{\sqrt{2^{\,n_3+n_4}}}
        \sum_{r=0}^{n_4}\sum_{s=0}^{n_3}
        \binom{n_4}{r}\binom{n_3}{s}
        (-1)^{\,n_3-s}
        \\[4pt]
         & \quad\times
        \sum_{n_5,n_6=0}^{\infty}
        \frac{(-1)^{\,n_6}\sqrt{P_{n_5}P_{n_6}}}{n_5!\,n_6!}
        \frac{1}{\sqrt{2^{\,n_5+n_6}}}
        \sum_{u=0}^{n_5}\sum_{v=0}^{n_6}
        \binom{n_5}{u}\binom{n_6}{v}
        (-1)^{\,n_6-v}
        \\[6pt]
         & \quad\times
        2^{-\frac{n_H+m_H+n_V+m_V}{2}}\;
        \delta_{k_2,\;n_H+m_H-i_2}\;
        \delta_{l_2,\;n_V+m_V-j_2}\;
        \mathcal{F}_H(n_H,m_H;i_2)\;
        \mathcal{F}_V(n_V,m_V;j_2)
        \\[6pt]
         & \quad\times
        (\hat a_{1H}^\dagger)^{n_1}\,
        (\hat a_{1V}^\dagger)^{n_2}\,
        (\hat b_{3H}^\dagger)^{n_6}\,
        (\hat b_{3V}^\dagger)^{n_5}\,
        \ket{0}.
    \end{aligned}
    \label{equation_A17}
\end{equation}

where

\begin{equation}
    n_H=r+s,\qquad
    n_V=n_3+n_4-r-s,\qquad
    m_H=u+v,\qquad
    m_V=n_5+n_6-u-v.
    \label{equation_A18}
\end{equation}

and

\begin{align}
    \mathcal{F}_H(n_H,m_H;i_2)
     & :=
    \sum_{x=\max(0,\,i_2-m_H)}^{\min(n_H,\,i_2)}
    \binom{n_H}{x}\binom{m_H}{i_2-x}\,
    (-1)^{\,m_H-i_2+x},
    \label{equation_A19}
    \\[6pt]
    \mathcal{F}_V(n_V,m_V;j_2)
     & :=
    \sum_{y=\max(0,\,j_2-m_V)}^{\min(n_V,\,j_2)}
    \binom{n_V}{y}\binom{m_V}{j_2-y}\,
    (-1)^{\,m_V-j_2+y}.
    \label{equation_A20}
\end{align}

The accepted records are grouped into
\(
S=\{1001,0110\},\qquad B=\{1100,0011\}.
\)
For split and same-rail Green Machine records, respectively,

\begin{equation}
\begin{split}
        \ket{\psi_S}
        =
        N\,\frac{P_{0}\sqrt{P_{0}}\,P_{1}}{4\sqrt{2}}\Big[
        &\sqrt{2 P_1}\big(\ket{1,0;0,1}_{a_1b_3}+(-1)^{m_1}\ket{0,1;1,0}_{a_1b_3}\big)\\
        &+(-1)^{m_2+m_3}\sqrt{P_{2}}\big(\ket{1,1;0,2}_{a_{1}b_{3}}-\ket{1,1;2,0}_{a_{1}b_{3}}\big)\Big],
    \label{equation_A21}
\end{split}
\end{equation}
\begin{equation}
\begin{split}
        \ket{\psi_B}
        =
        N\,\frac{P_{0}\sqrt{P_{0}}\,P_{1}}{4\sqrt{2}}\Big[
        &\sqrt{2 P_1}\big(\ket{1,0;1,0}_{a_1b_3}+(-1)^{m_1}\ket{0,1;0,1}_{a_1b_3}\big)\\
        &+(-1)^{m_2+m_3}\sqrt{P_{2}}\big(\ket{1,1;0,2}_{a_{1}b_{3}}-\ket{1,1;2,0}_{a_{1}b_{3}}\big)\Big],
    \label{equation_A22}
\end{split}
\end{equation}

The associated phase indices are listed in Table~\ref{tab:three_source_phase_indices}; the
normalization is $N^{2}=16/[P_{0}^{3}P_{1}^{2}(2P_{1}+P_{2})]$, so that $N^{-2}$
is the per-record generation probability $P_{\rm gen}^{\rm PBP}$ of
Eq.~\eqref{eq:Pgen3_lossless}.

\begin{table}[tb]
\centering
\begin{tabular}{c|cc@{\qquad}c|c}
\multicolumn{3}{c}{BSM} & \multicolumn{2}{c}{GM}\\
Record & $m_1$ & $m_2$ & Record & $m_3$\\
\hline
1001 & 1 & 1 & 1001 & 0\\
0110 & 1 & 0 & 0110 & 1\\
1100 & 0 & 1 & 1100 & 0\\
0011 & 0 & 0 & 0011 & 1
\end{tabular}
\caption{Phase indices $m_1$ and $m_2$ (BSM record) and $m_3$ (GM record) in
Eqs.~\eqref{equation_A21} and~\eqref{equation_A22}, verified numerically for all
16 record combinations.}
\label{tab:three_source_phase_indices}
\end{table}

If the Green Machine is replaced by a second conventional BSM, the state becomes

\begin{equation}
\begin{split}
\ket{\psi_{\rm BSM}} =\; & N'\frac{P_{0}\sqrt{P_{0}}\,P_{1}}{4\sqrt{2}}
\Bigg[\sqrt{2P_{1}}\left(\ket{1,0;0,1}_{a_{1}b_{3}}\pm\ket{0,1;1,0}_{a_{1}b_{3}}\right)\\
&\pm\left(\sqrt{2P_{0}}\ket{0,0;0,0}_{a_{1}b_{3}}\pm\sqrt{2P_{2}}\ket{1,1;1,1}_{a_{1}b_{3}}\right)\Bigg],
\end{split}
\label{eq:psiBSM3_app}
\end{equation}
with $N'^{2}=16/[P_{0}^{3}P_{1}^{2}(P_{0}+2P_{1}+P_{2})]$, so that
$N'^{-2}=P_{\rm gen}^{\rm BSM}$. The pre-postselection Bell fractions of the two
chains, $2P_1/(2P_1+P_2)=2G/(3G-1)$ and $2P_1/(P_0+2P_1+P_2)$, are
Eqs.~\eqref{eq:Fph3_lossless} and~\eqref{eq:Bph3_BSM_lossless} of the main text.
Projecting out the vacuum component on either side (ideal memory loading) removes
the $\ket{0,0;0,0}$ term of Eq.~\eqref{eq:psiBSM3_app} and leaves both chains
with the common postselected fidelity $2P_1/(2P_1+P_2)$ quoted after
Eq.~\eqref{eq:Bph3_BSM_lossless}.

\subsection{Four-source configuration}

The four-source architecture uses the same source state, BSM transformation, and Green Machine transformation introduced above. Sources 1 and 2 are conditioned by $\mathrm{BSM}_1$. Sources 3 and 4 are conditioned by $\mathrm{BSM}_2$, with
\(
n_5:(a_{3H},b_{3V}),\quad
n_6:(a_{3V},b_{3H}),\quad
n_7:(a_{4H},b_{4V}),\quad
n_8:(a_{4V},b_{4H}).
\)
The conditioned state of sources 3 and 4 is

\begin{equation}
    \begin{aligned}
        \ket{\psi'}_{34}
         & =
        \hat{\Pi}^{(i_2 j_2 k_2 l_2)} \, \hat{U}_{B_2} \ket{\psi}_{34}
        \\[4pt]
         & =
        \sqrt{i_2!\,j_2!\,k_2!\,l_2!}
        \sum_{\substack{n_5,n_6,n_7,n_8=0 \\ p_1,p_2,p_3,p_4}}^\infty
        \frac{(-1)^{\,n_6+n_8+p_1+p_3}}{n_5!\,n_6!\,n_7!\,n_8!}
        \frac{\sqrt{P_{n_5}P_{n_6}P_{n_7}P_{n_8}}}{\sqrt{2^{\,n_5+n_6+n_7+n_8}}}
        \\[4pt]
         & \quad\times
        \binom{n_6}{p_1}\binom{n_7}{p_2}\binom{n_5}{p_3}\binom{n_8}{p_4}
        \\[4pt]
         & \quad\times
        \delta_{i_2,\;p_1+p_2}\,
        \delta_{j_2,\;p_3+p_4}\,
        \delta_{k_2,\;n_6+n_7-p_1-p_2}\,
        \delta_{l_2,\;n_5+n_8-p_3-p_4}
        \\[4pt]
         & \quad\times
        (\hat a_{3H}^\dagger)^{n_5}
        (\hat a_{3V}^\dagger)^{n_6}
        (\hat b_{4H}^\dagger)^{n_8}
        (\hat b_{4V}^\dagger)^{n_7}
        \ket{0}.
    \end{aligned}
    \label{equation_A10}
\end{equation}

The two conditioned states are then combined using the same Green Machine transformation given in the three-source derivation. The resulting eight-index expression is obtained by replacing the standalone source-3 state by \(\ket{\psi'}_{34}\), so the repeated expansion is omitted.

For the representative case in which both partial BSM stages register 1001, the state before the final Green Machine projection is

\begin{equation}
    \begin{split}
        \ket{\psi}
        =
        \left(\frac{P_0P_1}{2}\right)^2 \frac{1}{2}
        \Bigg[
             \hat b_{2H'}^\dagger \hat a_{3V'}^\dagger
            \left(
            -\hat a_{1H}^\dagger \hat b_{4V}^\dagger
            + (-1)^{\mu_1+\nu_1}\hat a_{1V}^\dagger \hat b_{4H}^\dagger
            \right)
            +
            \hat b_{2V'}^\dagger \hat a_{3H'}^\dagger
            \left(
            \hat a_{1H}^\dagger \hat b_{4V}^\dagger
            - (-1)^{\mu_1+\nu_1}\hat a_{1V}^\dagger \hat b_{4H}^\dagger
            \right)
        \\
            +
            \hat b_{2H'}^\dagger \hat b_{2V'}^\dagger
            \left(
            -(-1)^{\nu_1}\hat a_{1H}^\dagger \hat b_{4H}^\dagger
            + (-1)^{\mu_1}\hat a_{1V}^\dagger \hat b_{4V}^\dagger
            \right)
            +
            \hat a_{3H'}^\dagger \hat a_{3V'}^\dagger
            \left(
            (-1)^{\nu_1}\hat a_{1H}^\dagger \hat b_{4H}^\dagger
            - (-1)^{\mu_1}\hat a_{1V}^\dagger \hat b_{4V}^\dagger
            \right)
            \Bigg]\ket{0}.
    \end{split}
    \label{equation28}
\end{equation}

For the Green Machine record 1001, the heralded outer state is
\(
\left(
            -\hat a_{1H}^\dagger \hat b_{4V}^\dagger
            + \hat a_{1V}^\dagger \hat b_{4H}^\dagger
            \right)\ket{0},
\)
which is the singlet state. The phase indices for the other accepted records are listed in Table~\ref{tab:four_source_phase_indices}, and the resulting record-to-state mapping is that of Table~\ref{tab:outer}.

\begin{table}[t]
    \centering
    \begin{tabular}{ccccc}
        \hline\hline
        Record & $\mu_1$ & $\mu_2$ & $\nu_1$ & $\nu_2$ \\
        \hline
        0011          & 0       & 0       & 0       & 0       \\
        1100          & 0       & 1       & 0       & 1       \\
        1001          & 1       & 1       & 1       & 1       \\
        0110          & 1       & 0       & 1       & 0       \\
        \hline\hline
    \end{tabular}
    \caption{Phase indices \(\mu_1\), \(\mu_2\), \(\nu_1\), and \(\nu_2\) associated with the four accepted records at each partial BSM.}
    \label{tab:four_source_phase_indices}
\end{table}

Thus, in the ideal lossless case, all accepted record combinations herald a pure Bell state between the outer modes.

\subsection{All-orders lossless extension}

The cancellation found for four sources extends to every alternating chain with
$N\geq4$ sources. The proof follows from photon-number conservation and does not
require truncating the SPDC state. Let $k_i$ denote the total number of pairs
emitted by source $i$. A Sagnac source places exactly $k_i$ photons in each of
its two spatial output rails. Measurement stage $j$ receives one rail from
source $j$ and one rail from source $j+1$. Because passive linear optics
conserves total photon number and every accepted PNR record contains exactly two
photons,
\begin{equation}
k_j+k_{j+1}=2,\qquad j=1,\ldots,N-1.
\label{eq:chain_number_constraint}
\end{equation}
It follows recursively that every accepted event has the pair-number history
\begin{equation}
(k_1,k_2,k_3,\ldots)
=(q,2-q,q,2-q,\ldots),
\qquad q\in\{0,1,2\}.
\label{eq:chain_number_histories}
\end{equation}
Thus no emission with more than two pairs from any source can contribute to a
lossless accepted record, even though the source state contains arbitrarily high
orders.

The $q=1$ history is the desired one-pair-per-source sector. Within this sector,
each accepted BSM or GM record is a maximally entangled two-qubit projection.
Contracting a chain of singlets with these projections leaves a maximally
entangled state on the two outer rails; equivalently, the associated $2\times2$
coefficient matrices are proportional to unitaries, and their product is also
proportional to a unitary. The records determine only the local correction
relating the output to a standard Bell state.

It remains to eliminate the $q=0$ and $q=2$ histories, which alternate vacuum
and double-pair sources. Up to normalization, the two-pair component of an
internal Sagnac source is
\begin{equation}
\begin{split}
\left(
\hat a_H^\dagger\hat b_V^\dagger
-\hat a_V^\dagger\hat b_H^\dagger
\right)^2
={}&(\hat a_H^\dagger)^2(\hat b_V^\dagger)^2
-2\hat a_H^\dagger\hat a_V^\dagger
  \hat b_H^\dagger\hat b_V^\dagger \\
&+(\hat a_V^\dagger)^2(\hat b_H^\dagger)^2 .
\end{split}
\label{eq:sagnac_two_pair_classes}
\end{equation}
The first and third terms have same-polarization double emissions on both rails
and are rejected by the BSM; the middle term has mixed-polarization double
emissions and is rejected by the GM. Hence an internal double-pair source lying
between a BSM and a GM is annihilated in its entirety. For the
$(0,2,0,2,\ldots)$ history, source 2 provides such an internal source. For the
$(2,0,2,0,\ldots)$ history, source 3 does so whenever $N\geq4$. The latter source
is an endpoint when $N=3$, which is why the residual four-photon term
survives in the minimal cascade. Therefore, under lossless propagation, ideal
PNR detection, and the stated two-photon acceptance rule, every alternating
BSM--GM chain with $N\geq4$ heralds a pure outer Bell state to all SPDC emission
orders.

As an independent check, we implemented the sequential Fock recurrence
\begin{equation}
\ket{\psi_j(\bm r)}=
\bra{r_j}U_j
\left(\ket{\psi_{j-1}(\bm r)}\otimes\ket{\mathrm{Sagnac}}_j\right),
\label{eq:finite_chain_recurrence}
\end{equation}
where $U_j$ alternates between the BSM and GM transformations and
$r_j\in\{1001,0110,1100,0011\}$. After each projection the four measured modes
are removed, so only the two outer modes and the next source need remain active.
We retained occupations through $n_{\max}=2$ in each two-mode-squeezed component.
Equation~\eqref{eq:chain_number_histories} shows that this truncation is exact for
the accepted lossless state: higher pair-number sectors cannot reach any accepted
record.

Table~\ref{tab:finite_chain_check} summarizes an exhaustive enumeration. Every
accepted record sequence has nonzero probability and leaves exactly two
equal-magnitude outer amplitudes with either $\psi$- or $\phi$-type support.
Thus every sequence heralds a Bell state up to a record-dependent local phase.
All $64$, $256$, and $1024$ accepted record sequences for $N=4$, $5$, and $6$
are nonzero and yield unit photonic fidelity, in agreement with the general
proof.
Loss breaks Eq.~\eqref{eq:chain_number_constraint}, because a higher-order
emission can lose photons before producing a two-photon record; the lossy case
therefore requires the all-orders Gaussian treatment used in the main text.
\begin{table}[tb]
\centering
\begin{tabular}{cccc}
\hline\hline
sources & analyzer sequence & record sequences & minimum $\calF_{\rm Ph}$ \\
\hline
$4$ & BSM--GM--BSM & $4^3=64$ & $1$ \\
$5$ & BSM--GM--BSM--GM & $4^4=256$ & $1$ \\
$6$ & BSM--GM--BSM--GM--BSM & $4^5=1024$ & $1$ \\
\hline\hline
\end{tabular}
\caption{Exhaustive check of the all-orders lossless result for finite chains.
The fidelity is minimized over all accepted record sequences.}
\label{tab:finite_chain_check}
\end{table}

\bibliography{ref2}
\end{document}